%% file: paper.tex
\documentclass[format=acmsmall, review=false, screen=true]{acmart}

\usepackage{booktabs} 

\acmJournal{TIIS}
\acmYear{2018}

\setcopyright{acmcopyright}
\acmJournal{TIIS}
\acmYear{2019} \acmVolume{1} \acmNumber{1} \acmArticle{1} \acmMonth{1} \acmPrice{15.00} \acmDOI{10.1145/3326128}

\usepackage{amssymb}
\usepackage{amsmath}
\usepackage{bm} 
\usepackage{xspace}
\usepackage{algorithm}
\usepackage{algpseudocode}
\usepackage{varwidth}
\usepackage{threeparttable}
\usepackage{multirow,bigdelim} 
\usepackage{dsfont}
\usepackage{epsfig}
\usepackage{graphicx}
\usepackage{hyperref}
\usepackage{xspace}
\usepackage{color}
\usepackage{epstopdf}
\usepackage{chngpage}
\usepackage{subcaption}
\usepackage{setspace}
\usepackage{bbm}
\usepackage{enumerate}

\usepackage{diagbox}

\newcommand*{\ML}{Movielens\xspace}
\newcommand*{\minimize}{\operatornamewithlimits{minimize}}

\newcommand*{\ARM}{ARM\xspace}
\newcommand*{\ES}{ESARM\xspace}
\newcommand*{\VO}{VOARM\xspace}
\newcommand*{\SETAVG}{SetAvg\xspace}
\newcommand*{\MFSET}{MFSET\xspace}
\newcommand*{\MFOPT}{MFOpt\xspace}
\newcommand*{\MLREALSETS}{ML-RealSets\xspace}
\newcommand*{\SYNSETSA}{SynSets\xspace}

\newcommand*{\upicky}{$U_{\mbox{\emph{\scriptsize Picky}}}$\xspace}
\newcommand*{\unonpicky}{$U_{\mbox{\emph{\scriptsize Non-picky}}}$\xspace}

\begin{document}

\title{Learning from Sets of Items in Recommender Systems}\thanks{This work was
supported in part by NSF (1447788, 1704074, 1757916, 1834251), Army
Research Office (W911NF1810344), Intel Corp, and the Digital Technology Center
at the
University of Minnesota. Access to research and computing facilities was
provided by
the Digital Technology Center and the Minnesota Supercomputing Institute.}

\author{Mohit Sharma}
\author{F.Maxwell Harper}
\author{George Karypis}
\affiliation{%
  \institution{University of Minnesota}
  \country{USA}
}

\begin{abstract}
Most of the existing recommender systems use the ratings provided by users on
individual items.
An additional source of preference information is to use the ratings that users provide on sets of items.
The advantages of using preferences on sets are two-fold. First, a
rating provided on a set conveys some preference information about each of the
set's items, which allows us to acquire a user's preferences for more items that
the number of ratings that the user provided.
Second, due to
privacy concerns, users may not be willing to reveal their preferences on
individual items explicitly but may be willing to provide a single rating to a
set of items, since it provides some level of information hiding. This paper
investigates two questions related to using set-level ratings in
recommender systems. First, how users' item-level ratings relate to their
set-level ratings. Second, how collaborative filtering-based models for
item-level rating prediction can take advantage of such set-level ratings. 
We have collected set-level ratings from active users of \ML
on sets of movies that they have rated in the past. 
Our analysis of these ratings shows that though the majority of the users
provide the average of the ratings on a set's constituent items as the rating on
the set, there exists a significant number of users that tend to
consistently either under- or over-rate the sets.
We have developed collaborative filtering-based
methods to explicitly model these user behaviors that can be used to recommend
items to users. 
Experiments on real data and on synthetic data that resembles the under- or
over-rating behavior in the real data, demonstrate that these models can recover the
overall characteristics of the underlying data and predict the user's ratings on
individual items.
\end{abstract}

%
%

%
%


\maketitle


\section{Introduction} \label{intro}
\input{intro.tex}

\section{Related Work} \label{related_work}

\input{relatedwork.tex}

\section{Movielens set ratings dataset} \label{dataset}
\input{dataset.tex}

\section{Methods} \label{lfs_method}
\input{method.tex}

\section{Experimental Evaluation} \label{exp_eval}
\input{experiments.tex}

\section{Results and Discussion} \label{results}
\input{results.tex}

\section{Conclusion and future work} \label{conclusion}
\input{conclusion.tex}

\bibliographystyle{ACM-Reference-Format-Journals}
\bibliography{refs}

\end{document}

%% file: intro.tex


Recommender systems help consumers by providing suggestions that are
expected to satisfy their tastes. They are successfully  deployed in several domains
such as e-commerce (e.g., Amazon, Ebay), multimedia content providers (e.g., Netflix,
Hulu) and mobile app stores (e.g., Apple, Google Play).
Collaborative filtering~\cite{r30,koren2009matrix} which takes advantage of users' past preferences
to suggest relevant items, is one of the key methods
used by recommender systems.

Most collaborative filtering approaches rely on past preferences provided by
users on individual items.
An additional source of preferences is the user's preferences on sets of items.
Example of such set-level ratings includes
ratings on song playlists, music albums, reading lists, watchlists, vacation packages, product assortments, etc. 
A rating provided by the user on a set of items conveys some information about the
user's preference on each of the set's items and, as a result, it is a
mechanism by which some information about user's preferences can be acquired for
many items.
At the same time, due to privacy concerns, users that are not willing to
explicitly reveal their true preferences on individual items may provide a single
rating to a set of items, as it provides some level of information hiding.

This paper investigates two questions related to using set-level preferences
in recommender systems. 
First, how users' item-level ratings relate to the
ratings that they provide on a set of items. Second, how collaborative
filtering-based methods
can take advantage of such set-level ratings towards making item-level rating
predictions.

To answer the first question, we collected ratings on sets
of movies from users of \ML, a popular online movie recommender
system\footnote{www.movielens.org}. 
Our analysis of these ratings leads to two key findings. First, for the majority
of the users, the rating provided on a set can be accurately approximated by the
average rating that they provided on the set's constituent items. Second, there
is a considerable user population that tends to consistently either over- or
under-rate the set, especially for sets that contain items on which the user's
item-level ratings are diverse.
Using these insights, we developed different  models that can predict
a user's rating on a set of items as well as on individual items. 
Furthermore, these methods can use ratings on both the sets and the items and
lead to better results for the users that have either both or only one type of
ratings.
These methods solve
these problems in a coupled fashion by estimating models to predict the item-level
ratings and by estimating models that combine these individual ratings to derive
set-level ratings.

The key contributions of the work are the following:
\begin{enumerate}[(i)]
\item introduction of \emph{Variance Offset Average Rating Model} (\VO) and
\emph{Extremal Subset Average Rating Model} (\ES) to
model a user's consistency to over- or under-rate the set of items as a function of his/her ratings on the set's constituent items;
\item development of collaborative filtering-based methods that take
advantage of \VO and \ES in order to estimate users' preferences on sets of items as well as
on individual items; and
\item collection and analysis of a dataset that contains users' ratings both on
individual items and on various sets containing these items.
\end{enumerate}


This work significantly extends upon the preliminary work published earlier~\cite{sharmalearning} by expanding the analysis of the set-based ratings, by introducing a new approach to estimate the ratings from the set-level ratings (\ES), and by expanding the experimental evaluation.

The rest of the paper is organized as follows. Section~\ref{related_work} describes the relevant
prior work. Section~\ref{dataset} describes the dataset creation process
along with the analysis of the set ratings in relation to the users' ratings on
their constituent items. 
Section~\ref{lfs_method} presents the methods that we developed to estimate the
item-level models from the set ratings. 
Section~\ref{exp_eval} provides information about the evaluation methodology. 
Section~\ref{results} presents the results of the experimental evaluation. Finally, 
Section~\ref{conclusion} provides some concluding remarks.

%% file: relatedwork.tex
There has been little published work on using set-level ratings to improve the
accuracy of item-level recommendations. The one exception is a recent study in which relative
preference information on different groups of items was collected during a new user
signup process and these preferences were then used to assign a user to a set of
pre-built recommendation profiles~\cite{r53}. 
This approach significantly reduced the time required to learn the user's
preferences in order to generate recommendations for the new user.
The principal difference from this 
approach is that in our work we try to model the user behavior that determines
his/her estimated rating on a set and then use that to develop fully
personalized recommendation methods that are not limited to new users.

Another relevant problem is of energy disaggregation~\cite{hart1992nonintrusive}, which refers to the task of separating the energy signal of a building into the energy signals of
individual appliances that reside in the building. Disaggregated energy
consumptions are used to provide feedback to consumers,  forecast demands,
design energy incentives and detect appliances'
malfunction~\cite{froehlich2011disaggregated,darby2006effectiveness}. Similar to
the idea of energy disaggregation, in our work, we try to separate a user's rating on a set of
items into the users' ratings on items in the set and generate item
recommendations for the user.

Sets of items have also been used to investigate different interfaces~\cite{mcnee2003interfaces} and strategies~\cite{rubens2015active, Rashid02IUI} for preference elicitation in order to learn more about the users in  recommender systems. Some of these techniques~\cite{rubens2015active, Rashid02IUI} are designed to identify a set of items  for which item-level ratings are then elicited by the users. Though those approaches do use sets of items, their use is not related to how they are used in the methods that we develop and study in our work. Our work requires users to provide a single rating to the set and not to its individual items.

The researchers have also investigated how different aspects, e.g., rating
questions~\cite{bellogin2014magic}, reference
points~\cite{adomavicius2011recommender,cosley2003seeing,nguyen2013rating}, and
contextual factors~\cite{Winoto2010RUM}, can influence a user when
elicited to provide a rating on an item. In our work, we have investigated how
does the user provides a rating on a set of items and used the derive insights to develop
collaborative filtering-based methods to predict the rating for an individual
item in the set.

In addition, there has been some work that  has focused on recommending
lists of items or bundles of items. For example, recommendation of music
playlists~\cite{r55,moore2012learning,aizenberg2012build}, travel
packages~\cite{interdonato2013versatile,r54,liu2011personalized,xie2011comprec}, reading
lists~\cite{r56} and recommendation of lists under user specified budget
constraints~\cite{xie2010breaking,BenouaretRecsys16}.
However, this research is not directly related to the problems explored in this
paper because our focus is on learning the user's ratings on items in lists from
the ratings that the user provided on these lists.

Matrix factorization (MF) is one of the widely used collaborative
filtering-based methods 
in recommender systems~\cite{hu2008collaborative,koren2008factorization,koren2010collaborative,koren2009matrix}. 
The  MF method assume that the
user-item rating matrix is low-rank and can be computed as a product of two
matrices known as the user and the item latent factors. 
If for user $u$, the vector $\bm{p_u} \in \mathbb{R}^f$ denotes the $f$ dimensional
user's latent representation and similarly for item $i$, the vector 
$\bm{q_i} \in \mathbb{R}^f$ represents the $f$ dimensional item's latent
representation, then the
predicted rating for user $u$ on item $i$, i.e., $\hat{r}_{ui}$ is given by
\begin{equation} \label{mf_eq}
  \begin{split}
    \hat{r}_{ui} = \bm{p_u}^T\bm{q_i}.
  \end{split}
\end{equation}

The user and item latent factors are learned by minimizing a regularized
square loss between the actual and predicted ratings 
\begin{equation} \label{mf_obj}
  \begin{aligned}
    \operatornamewithlimits{minimize}_{P, Q} & &\frac{1}{2}\sum_{r_{ui} \in R}
    \left(r_{ui} - \bm{p}_u^T \bm{q}_i \right)^2+ \frac{\beta}{2}
    \left(||P||_F^2 + ||Q||_F^2 \right),
  \end{aligned}
\end{equation}
where the matrices $P \in \mathbb{R}^{m \times f}$ and $Q \in\mathbb{R}^{n \times f}$ 
contain the latent factors of the users and the items
respectively. The parameter $\beta$ controls the Frobenius norm regularization
of the latent factors to prevent overfitting. 
Equation~\ref{mf_obj} can be solved by using Stochastic Gradient Descent (SGD)
method~\cite{koren2009matrix}.

%% file: dataset.tex

In order to study how the rating a user provides to a set of items relates to the ratings that the user provides on the individual items, we built a system to collect such set-level ratings and analyzed the data that were collected. The system that we developed and the analysis that we performed are described in the rest of this section. Specifically, Section~\ref{data_collection}  and Section~\ref{data_processing} describe the data collection and data pre-processing steps, respectively.
Section~\ref{data_analysis}, investigates (i) if the collected ratings are distributed uniformly or if some ratings tend to appear more than others, (ii) how a user's rating on a set relates to the user's ratings on individual movies, (iii) if the diversity of the ratings of the movies in a set could lead a user to under- or over-rate the set, (iv) whether the recently rated items carry more weight than the items rated a long time ago, (v) if the difference in the content of the items in a set could lead a user to under- or over-rate the set; and (vi) if there are users that tend to consistently under- or over-rate sets.

\subsection{Data collection}\label{data_collection}
\ML is a recommender system that utilizes
collaborative filtering algorithms to recommend movies to their users based on
their preferences. We developed a set rating widget to obtain ratings on a set
of movies from the \ML users.  
The set rating widget could be
rated from 0.5 to 5 with a precision of 0.5.  
For the purpose of data collection, we selected users who were active since
January 2015 and have rated at least 25 movies. The selected users were
encouraged to participate by contacting them via email. 
The sets of movies that we asked a user to rate were created by selecting five
movies at random without replacement from the movies that they have already
rated. 
Hence, the user was familiar with the movies in the set that we asked him/her to rate.
Furthermore, we limited the number of
sets a user can rate in a session to 50, though users can potentially rate
more sets in different sessions.
The set rating widget went live on February 2016 and, for the purpose of this
study, we used the set ratings that were provided until April 2016.

\begin{figure}[ht]
  \centerline{\includegraphics[scale=0.28]{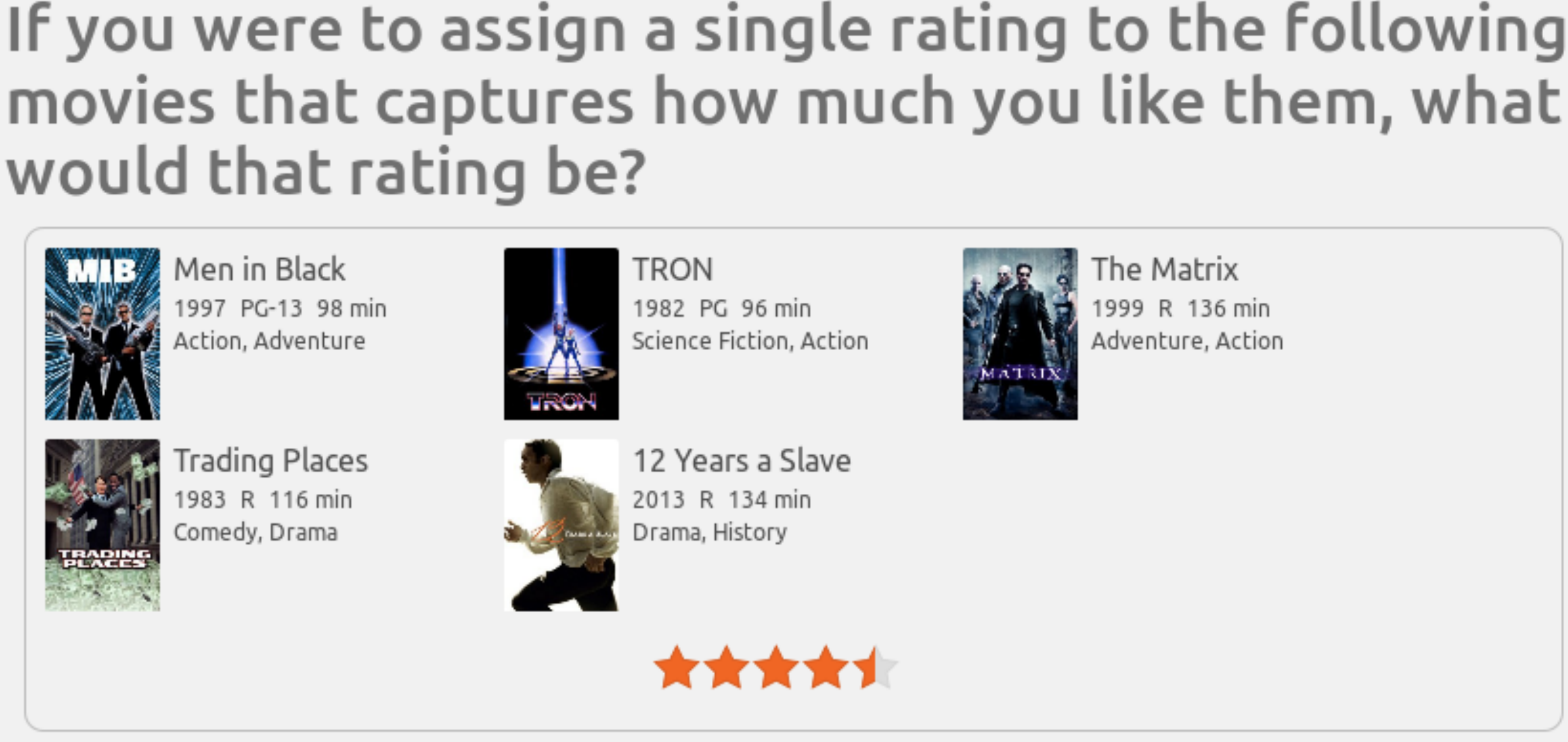}}
  \caption{The interface used to elicit users' ratings on a set of movies.}
  \label{fig:mlset}
\end{figure}

\subsection {Data processing}\label{data_processing}
\begin{figure}[bt]
  \centerline{\includegraphics[scale=0.9]{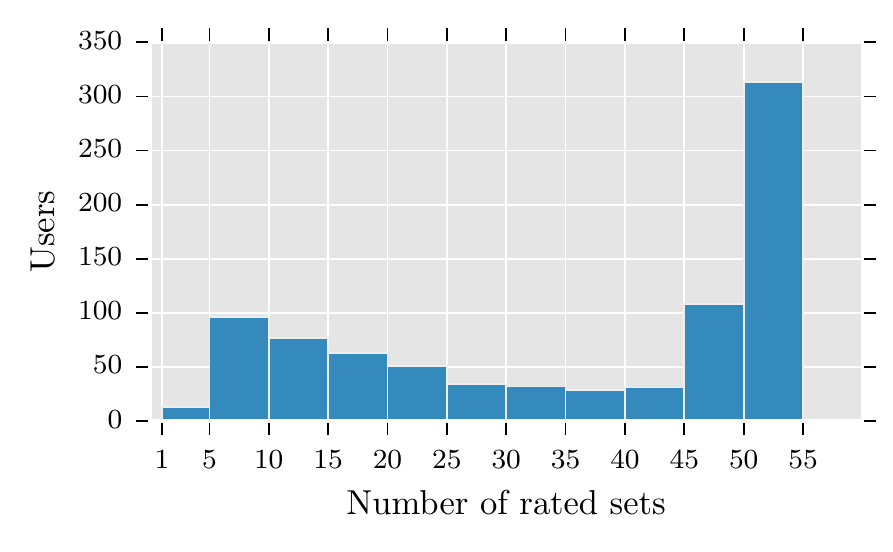}}
  \caption{The distribution of number of sets rated by the users.}
  \label{fig:usersetdist}
\end{figure}

From the initially collected data, we removed users who have rated sets within a
time interval of less than one second to avoid users who might be providing
the ratings at random. After this pre-processing, we were left with ratings
from 854 users over 29,516 sets containing 12,549 movies.
This dataset, after pre-processing, is available publicly to the community for further research.\footnote{\url{https://grouplens.org/datasets/learning-from-sets-of-items-2019/}}
Figure~\ref{fig:usersetdist} 
shows the distribution of the number of sets rated by the users,
which shows that roughly half of the users have rated at least 45 sets in a session.

\subsection{Analysis of the set ratings}\label{data_analysis}

\begin{figure}[t]
  \centerline{\includegraphics[scale=0.82]{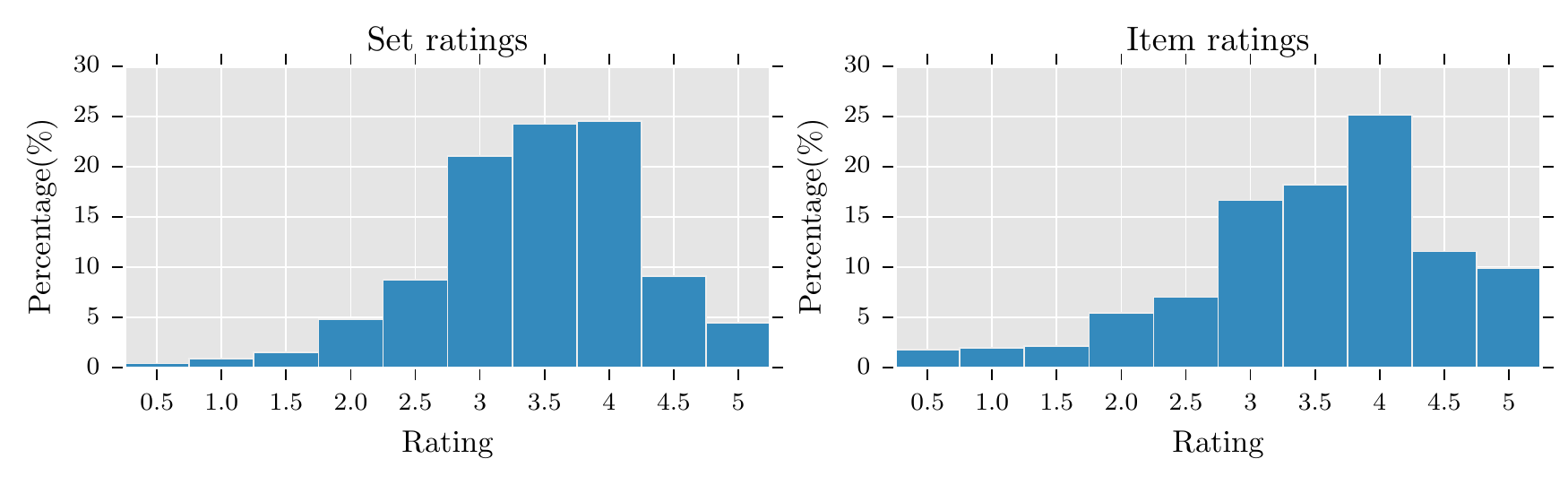}}
  \caption{The distribution of the provided set ratings (left) and the ratings of
  their constituent items (right).}
  \label{fig:itemsetratingdist}
\end{figure}

We investigated whether ratings are distributed uniformly or if some ratings
tend to appear more than others.
Figure~\ref{fig:itemsetratingdist} (left) depicts
the distribution of the collected ratings over all the sets. 
The majority of the ratings lie between 3.0 and 4.0. 
Since, by construction, we know the actual ratings that these users provided on
the actual movies. Figure~\ref{fig:itemsetratingdist} (right) shows the
distribution of the ratings of the movies that were contained in all these sets.
By comparing these distributions we can see that the average item-rating (3.50) is
somewhat higher than the average set-based rating (3.44) but the overall variance of
the set-based ratings (0.65) is lower than that of the item ratings (1.01).

\begin{figure}[tb]
  \centerline{\includegraphics[scale=0.82]{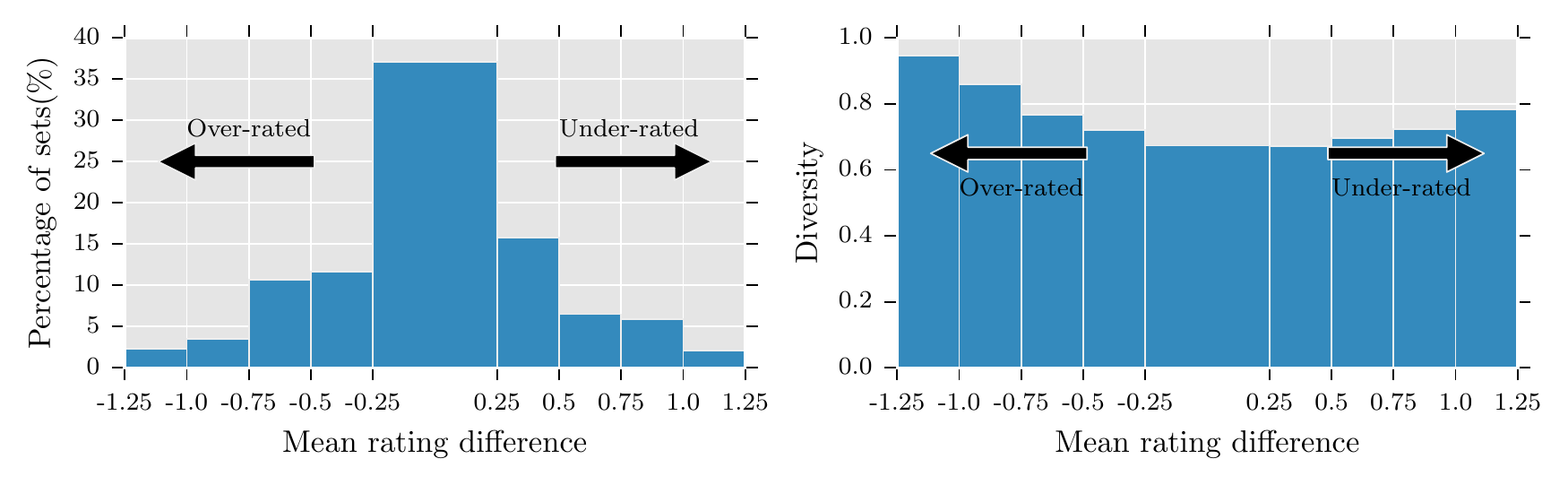}}
  \caption{Histogram of percentage of sets (left) and diversity (right) against
  mean rating difference (MRD).}
  \label{fig:fractiondiversitymonths}
\end{figure}

In order to analyze how consistent a user's rating on a set is with the ratings provided by
the user on the movies in the set, 
we computed the difference of the average of the user's  ratings on the items in the set
and the rating assigned by a user to the set. We
will refer to this difference as \emph{mean rating difference} (MRD). 
Figure~\ref{fig:fractiondiversitymonths} (left) shows the distribution of the MRD values in
our datasets. The majority of the sets have
an MRD within a margin of 0.5 indicating that the users have rated them close to the
average of their ratings on set's items. The remaining of the sets have been
rated either significantly lower or higher from the average rating. We refer 
to these sets as the under- and
the over-rated sets, respectively. Moreover, an interesting observation from the
results in Figure~\ref{fig:fractiondiversitymonths} (left), is that the number of
under-rated sets is more than that of the over-rated sets.

\begin{figure}[bt]
  \centerline{\includegraphics[scale=0.7]{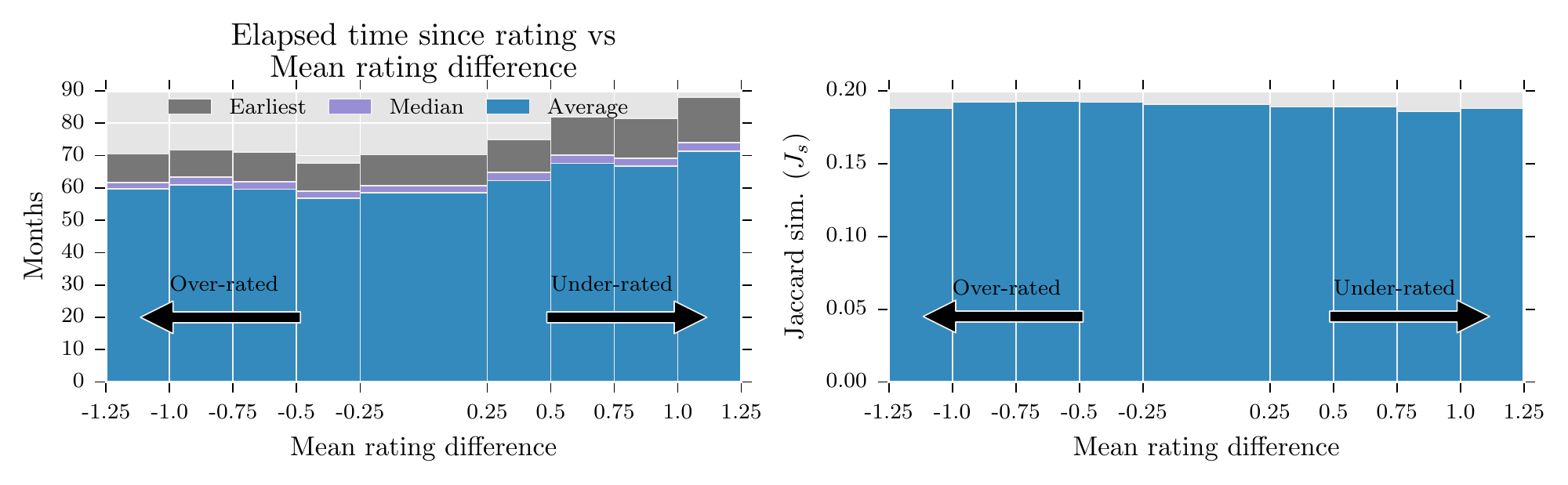}}
  \caption{Histogram of elapsed time in months (left) and jaccard similarity of movies (right) against mean rating difference (MRD).}
  \label{fig:elapsedmonths_jaccsim}
\end{figure}

\begin{figure*}[tb]
  \centerline{\includegraphics[scale=0.82]{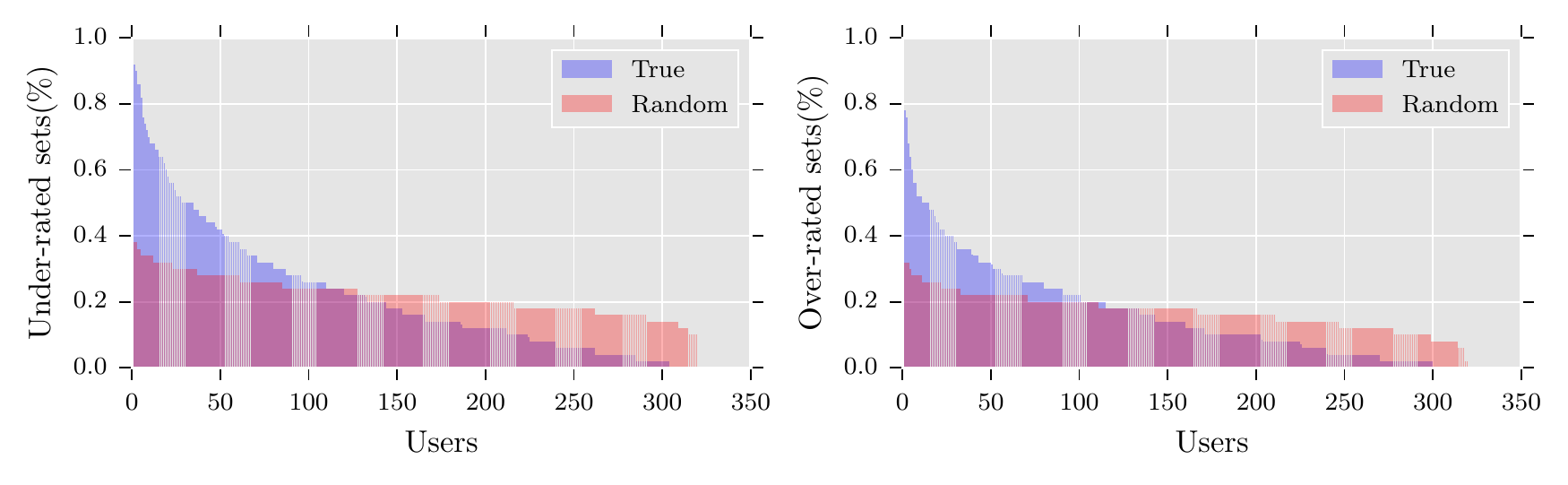}}
  \caption{Fraction of under-rated and over-rated sets across users in true and
  random population.}
  \label{fig:underrated}
\end{figure*}

In order to understand what can lead to a set being under-
or over-rated, we investigated if the \emph{diversity} of the ratings of the
individual movies in a set could lead a user to under- or over-rate the set.
We measured the diversity of a set as the standard deviation of the ratings
that a user has provided to the individual items of the set.
As shown in Figure~\ref{fig:fractiondiversitymonths} (right), the sets 
that contain more diverse ratings (i.e., higher standard deviations) tend
to get under- or over-rated more often when compared to less diverse sets. This
trend was found to be
statistically significant ($p$-value of 0.01 using $t$-test).

Furthermore, we investigated whether the recently rated items
carry more weight than the items rated a long time ago. To this end, we computed
the difference between the timestamp of the earliest rating of the movies in the
set and the year 2016, i.e., when the users were asked to rate the sets.
Similarly, we computed the median and average age of movies in a set.
Interestingly as shown in Figure~\ref{fig:elapsedmonths_jaccsim} (left), the under-rated sets
contained movies whose ratings were provided on average five
years before the survey while the remaining sets contained the movies whose
ratings were provided on average four years before the survey. This difference
among the sets was found to be statistically significant ($p$-value $<$ 0.001
using $t$-test). 
This suggests that the user's preference for a movie rated in the past carries
lower weight than the recently rated movie. 
The user's higher preference for a recent movie is not surprising as it has been
shown that the user tends to rate a movie close to the middle of the scale as
the time between viewing a movie and rating it increases~\cite{bollen2012remembering}.

Moreover, the difference in the content of the items in a set may also lead a user to under- or over-rate the set. We examined if the difference in genres of movies in a set can lead to under- and over-rating of the set. To this end, we computed average pairwise \emph{jaccard} similarity of the movies in a set after considering genres of the movies in the set. 
Average pairwise jaccard similarity ($J_s$) of the movies in set $\mathcal{S}$ is given by
\begin{equation} \label{avgJSimEq}
  \begin{split}
    J_{s} = \frac{2}{|\mathcal{S}| (|\mathcal{S}| -1)} \sum_{i=1}^{|\mathcal{S}|} \sum_{j = i+1}^{|\mathcal{S}|} \frac{|\mathcal{G}_i \cap \mathcal{G}_j|}{|\mathcal{G}_i \cup \mathcal{G}_j|},
  \end{split}
\end{equation}
\noindent where $|\mathcal{S}|$ denotes size of set $\mathcal{S}$ and $\mathcal{G}_k$ represents the set of genres of movie $k$ in set $\mathcal{S}$.
Interestingly, as can be seen in Figure~\ref{fig:elapsedmonths_jaccsim} (right), the average jaccard similarity of movies in sets is comparable across the under- or over-rated sets and the variation of jaccard similarity was found not to be statistically significant  ($p$-value of  0.769 using $t$-test). 
The insignificant variation in jaccard similarity suggests that a user rating on a set of movies is not influenced by the difference in genres of the movies in the datset.

Additionally, we studied if there are users that tend to consistently under- or
over-rate sets. To this end, we selected users who have rated at
least 50 sets and computed the fraction of their under- and over-rated sets.
We also computed the fraction of under- and over-rated sets across a random
population of the same size. We generated this random population by randomly
permuting the under-rated and over-rated sets across the users.
Figure~\ref{fig:underrated} shows the fraction of under- and over-rated sets
for both the true and random population of user. 
In the true population, some users tend to under- or over-rate sets
significantly more than that of the random population. Using the Kolmogorov-Smirnov 2 sample test, 
we found this behavior of true population to be statistically different ($p$-value $<$ 0.001) from that of 
random population. 

The above analysis reveals that our dataset contains users that when they are
asked to assign a single rating to a set of
items, some of them consistently assign a rating that is lower than the
average of the ratings that they provided
to the set's constituent items (they under-rate), whereas others assign a rating
that is higher (they over-rate). Thus some
users are very demanding (or picky) and tend to focus on the worst items in the set,
whereas other users are less demanding and
tend to focus on the best items in the set. 
Henceforth, we will refer to the tendency of a user to under-rate sets of items as the user-specific~\emph{pickiness}.
Moreover, we will refer to a user as being picky if the user under-rates a set and less picky if the user over-rates the set.

%% file: method.tex

In this section, we investigate different approaches that capture the user behavior of providing ratings on sets. We describe various methods that use the set ratings alone or in combination with individual item ratings towards solving two problems: (i) predict a rating for a set of items, and (ii) predict a rating for individual items.  
Our methods solve these problems in a coupled fashion by estimating models for predicting the ratings that users will provide to the individual items and by estimating models that use these item-level ratings to derive set-level ratings.

\subsection{Modeling users' ratings on sets}
In order to estimate the preferences on individual items from the preferences on the sets, we need to make some assumptions on how a user derives a set-level
rating from the ratings of the set's constituent items.
Informed by our analysis of the data described in Section~\ref{dataset}, we
investigated three approaches of modeling that.

\subsubsection*{Average Rating Model (ARM)}
The first approach assumes that the rating that a user provides
on a set reflects his/her average rating on all the items in the
set. Specifically, if the rating of user $u$ on set $\mathcal{S}$ is denoted by $r_{u}^s$,
then the estimated rating of user $u$ on set $\mathcal{S}$ is given by
\begin{equation} \label{avgSetEq}
  \begin{split}
    \hat{r}_{u}^s &= \frac{1}{|\mathcal{S}|} \sum_{i \in \mathcal{S}} r_{ui}.
  \end{split}
\end{equation}
\noindent As the analysis in Section~\ref{dataset} showed, such a model correlates well
with the actual ratings that the users provided on majority of the sets,
especially when the ratings of the constituent items are not very different.

\subsubsection*{Extremal Subset Average Rating Model (ESARM)}
In order to capture the user-specific \emph{pickiness}, i.e., the tendency of a user to under-rate sets of items, illustrated in Figures~\ref{fig:fractiondiversitymonths}~and~\ref{fig:underrated}, 
this approach postulates that a user rates a set by considering only a subset of the set's items. If a user tends to
consistently under-rate each set, then that subset will contain some of each
set's lowest-rated items. Analogously, if a
user tends to consistently over-rate each set, then that subset will contain
some of each set's highest-rated items.
Moreover, this approach further postulates that given such subsets, the rating
that a user will assign to the set as a
whole will be the average of his/her ratings on the individual items of the
subset. The parameter in this model that
captures the level of a user's pickiness is the size of the subset and whether
or not it will contain the least- or the
highest-rated items. We will call these subsets having least- and highest- rated
items as \emph{extremal subsets}. The set rating of an extremely picky user will be
determined by the average rating of one or two of
the least rated items, whereas the set rating of a user that is not picky at all
will be determined by the average rating of
one or two of the highest rated items. 

If $e_i$ denotes the average rating of items in $i$th extremal
subset and $n_s$ denotes number of items in set $\mathcal{S}$, 
then $\langle e_1, \ldots, e_{n_s}, \ldots, e_{2n_s - 1} \rangle$ represents
the average rating on all the extremal subsets; for $1 \le i \le
n_s$, $e_i$ is the average rating of $i$ least rated
items, for $n_s \le i \le 2n_s - 1 $, $e_i$ is the average rating
of the $2n_s - i$ highest rated items and $e_{n_s}$ is the average rating of all
the items in the set. Then $\hat{r}_{u}^s$ is given by
\begin{equation} \label{avgbuckacteq}
  \begin{split}
    \hat{r}_{u}^s &= \sum_{i=1}^{2n_s - 1} w_{u,i} e_{i},
  \end{split}
\end{equation}
\noindent where $w_{u,i}$ is a non-negative weight of user $u$ on
$i$th extremal subset and the weights sum to 1.
The weight $w_{u,i}$ measures the influence of the items in $i$th extremal subset 
towards estimating the user's rating on set $\mathcal{S}$.
One of the weights corresponds to the extremal subset that is responsible for majority of the user's rating on set, and it is higher than others, i.e.,
\begin{equation} \label{esqpmodel_eq}
	\begin{aligned}  
        & \sum_{i=1}^{2n_s - 1}w_{u,i}=1,  \\
        & w_{u, j} < w_{u, j+1}, \forall j < k, \\
        & w_{u, j+1} < w_{u, j}, \forall j \ge k, \\
        & w_{u,k} > c,  c > 0, \\
     \end{aligned}
\end{equation}
\noindent where $c$ is the minimum weight of the extremal subset having the
    highest contribution towards the user's rating on set.

Note that this model assumes
that the size of all the sets are the same, however it can be generalized to sets
of different sizes by constructing the extremal subsets for fixed number
of quantiles in a set.

\subsubsection*{Variance Offset Average Rating Model (VOARM)}
This approach captures the user-specific \emph{pickiness} by assuming 
that a user rates a set by considering both the
average rating of the items in the set
and also the diversity of the set's items. In this model, the set's rating is
determined as the sum of the average
rating of the set's items and a quantity that depends on the sets diversity
(e.g.,
the standard deviation of the set's ratings) and the user's level of
pickiness. 
If a user is very picky, that quantity will be
negative and large, resulting to the set being
(severely) under-rated. On the other hand, if a user is not picky at all, that
quantity will be positive and large, resulting to the 
set being (severely) over-rated.

If $\beta_u$ denotes the pickiness level of user $u$,
then the estimated rating on a set is given by
\begin{equation} \label{varActEq}
  \begin{split}
    \hat{r}_{u}^s &= \mu_{s} + \beta_u \sigma_{s},
  \end{split}
\end{equation}
\noindent where $\mu_{s}$ and $\sigma_{s}$ are the mean and the standard
deviation of the ratings of items in the set $\mathcal{S}$.  
Both $\mu_{s}$ and $\sigma_{s}$ are given by 
\begin{equation}
  \mu_{s} = \frac{1}{|S|} \sum_{i \in \mathcal{S}} r_{ui},\;\;  
  \sigma_{s} = \sqrt{\frac{1}{|S|} \sum_{i \in \mathcal{S}} (r_{ui} -\mu_{s})^2}.
\end{equation}
\noindent In contrast to \ES, this method considers all the items in a set by considering the average rating of items in the set, i.e., $\mu_{s}$,  and the user's level of pickiness, i.e., $\beta_u$, determines how a user's rating on the set is affected by the least rated movies or the highest rated movies in the set. 

\subsection{Modeling user's ratings on items}
In order to model a users' ratings on the items, similar to matrix
factorization method~\cite{koren2009matrix},
we assume that the underlying user-item rating matrix is low-rank, i.e., there is a low-dimensional latent space in which both the users
and the items can be compared to each other. 
The rating of user $u$ on item $i$ can be computed as an inner product of
the user and the item latent factors in that latent space.
Thus, the estimated
rating of user $u$ on item $i$, i.e., $\hat{r}_{ui}$, is given by
\begin{equation} \label{ratPred_eq}
  \begin{split}
    \hat{r}_{ui} = \bm{p_u}^T\bm{q_i},
  \end{split}
\end{equation}
\noindent where $\bm{p_u} \in \mathbb{R}^f$  is the latent representation of user $u$,
$\bm{q_i} \in \mathbb{R}^f$ is the latent representation of item $i$ and $f$ is the
dimensionality of the underlying latent space.

\subsection{Combining set and item models}

Our goal is to estimate the item-level ratings by learning the user and item
latent factors of Equation~\ref{ratPred_eq}; however, the ratings that we have available
from the users are at the set-level. In order to use the available set-level
ratings, we need to combine 
Equation~\ref{ratPred_eq} with
Equations~\ref{avgSetEq},~\ref{avgbuckacteq}~and~\ref{varActEq}. To solve the problem, we assume that the actual
item-level ratings used in Equations~\ref{avgSetEq},~\ref{avgbuckacteq}~and~\ref{varActEq} correspond to the estimated ratings
given by Equation~\ref{ratPred_eq}. Hence, the estimated set-level ratings in
Equations~\ref{avgSetEq},~\ref{avgbuckacteq}~and~\ref{varActEq}
are finally expressed in terms of the corresponding user and item latent
factors.

\begin{algorithm}
  \caption{Learn ARM}
  \label{alg:alg-lfs-arm}
  \small
  \begin{algorithmic}[1]
    \Procedure{LearnARM}{}
    \State $\eta \gets  \text{learning rate}$
    \State $\lambda \gets \text{regularization parameter}$
    \State $\mathcal{R}^s \gets \text{all users' ratings on sets}$  
    \State $iter \gets 0$
    \State Init $P$, $Q$ with random values $\in$ [0,1] 
    \While {iter $<$ maxIter and RMSE on validation set decreases}
      \State $\mathcal{R}^s \gets \text{shuffle}(\mathcal{R}^s)$
      \ForAll{$r_{u}^s \in \mathcal{R}^s$}
        \State $\hat{r}_{u}^s \gets  \frac{1}{|\mathcal{S}|} \sum_{\substack{i
        \in \mathcal{S}}} \bm{p}_u^T\bm{q}_i$ 
        \State $e_{u}^s \gets (\hat{r}_{u}^s - r_{u}^s)$
        \State $\bm{v}_k \in \mathbb{R}^k$  $\gets 0$
        \For {each item $i \in s$ }
        \State $\bm{v}_k \gets \bm{v}_k + \bm{q}_i$
        \EndFor
        \State $\bm{p}_u \gets \bm{p}_u - \eta(\frac{e_{u}^s}{|s|} \bm{v}_k +
        \lambda \bm{p}_u)$ \Comment{Update user's latent representation}
        \For {each item $i \in s$ }
          \State $\bm{q}_i \gets \bm{q}_i - \eta(\frac{e_{u}^s}{|s|} \bm{p}_u +
            \lambda \bm{q}_i)$ \Comment{Update item's latent representation} 
        \EndFor
      \EndFor
      \State $iter \gets iter + 1$
    \EndWhile
    \EndProcedure
  \end{algorithmic}
\end{algorithm}

\begin{algorithm}
  \caption{Learn \ES}
  \label{alg:alg-lfs-esarm}
  \small
  \begin{algorithmic}[1]
    \Procedure{Learn\ES}{}
    \State $\eta \gets  \text{learning rate}$
    \State $\lambda \gets \text{regularization parameter}$
    \State $\mathcal{R}^s \gets \text{all users' ratings on sets}$  
    \State $n_s \gets \text{number of items in set}$
    \State $n_{es} \gets 2n_s - 1$ \Comment{number of possible extremal subsets}
    \State $iter \gets 0$
    \State Init $P$, $Q$ with random values $\in$ [0,1] 
    \State Init $W$ with random values $\forall$ user $u \in U$, s.t.,
    $\sum_{i=1}^{n_{es}}w_{u,i}=1$

    \While {iter $<$ maxIter and RMSE on validation set decreases}
      \State $\mathcal{R}^s \gets \text{shuffle}(\mathcal{R}^s)$
      \State 
      \ForAll{$r_{u}^s \in \mathcal{R}^s$}
        
        \State $\hat{r}_{u}^s \gets 0$
        \State $\mathcal{E}_s \gets \text{All possible extremal subsets
        for set }\mathcal{S}$
        \State         
        \State $\nabla p_u \in \mathbb{R}^f \gets 0$ 
        
        \For {each subset $i \in \mathcal{E}_s$}
          \State $\hat{e}_i \gets 0,\; q_{sum} \in \mathbb{R}^f \gets 0$
          \For {each item $j \in i$}
            \State $\hat{e}_i \gets \hat{e}_i + p_u^Tq_j, \; q_{sum} \gets q_{sum} + q_j$
          \EndFor
          \State $\hat{e}_i \gets \frac{\hat{e}_i}{|i|},\; q_{sum} \gets \frac{q_{sum}}{|i|}$
          \State $\hat{r}_{u}^s \gets \hat{r}_{u}^s + w_{u,i}\hat{e}_i$
          \State $\nabla p_u \gets \nabla p_u + w_{u,i}q_{sum}$
        \EndFor
        
        \State $e_{u}^s \gets (\hat{r}_{u}^s - r_{u}^s)$
        \State $\nabla p_u \gets 2e_{u}^s\nabla p_u +
        2\lambda p_u$ \Comment{update user's latent representation}
 
        \State $p_u \gets p_u - \eta \nabla p_u$
        \State         
        \State $\nabla q \gets 2e_u^s p_u$
        \For {each subset $i \in \mathcal{E}_s$}
          \For {each item $j \in i$}
          \State $q_j \gets q_j - \eta (\frac{w_{u,i}{\nabla q} }{|i|} +
            2\frac{\lambda}{n_s}q_j)$ \Comment{update items' latent representation} 
          \EndFor  
        \EndFor

      \EndFor
      
      \State
      \ForAll{$u \in U$}
        \State \text{Update } $\bm{w}_u$ \text{ using constraint quadratic
      programming as described} 
        \State \text{in Section}~\ref{model_learn}. 
      \EndFor

      \State
      \State $iter \gets iter + 1$
    \EndWhile

    \EndProcedure
  \end{algorithmic}
\end{algorithm}

\begin{algorithm}
  \caption{Learn \VO}
  \label{alg:alg-lfs-voarm}
  \small
  \begin{algorithmic}[1]
    \Procedure{Learn\VO}{}
    \State $\eta \gets  \text{learning rate}$
    \State $\lambda \gets \text{regularization parameter}$
    \State $\mathcal{R}^s \gets \text{all users' ratings on sets}$  
    \State $iter \gets 0$
    \State Init $P$, $Q$ and $\beta$s with random values $\in$ [0,1]
    \While {iter $<$ maxIter and RMSE on validation set decreases}
      \State $\mathcal{R}^s \gets \text{shuffle}(\mathcal{R}^s)$
      \ForAll{$r_{u}^s \in \mathcal{R}^s$}
        \State $\hat{\mu}_s \gets  \frac{1}{|\mathcal{S}|} \sum_{i \in \mathcal{S}} \bm{p}_u^T\bm{q}_i$ 
        \State $\hat{\sigma}_s \gets  \epsilon + \sqrt{\frac{1}{|\mathcal{S}|} \sum_{i \in
          \mathcal{S}}  (\bm{p}_u^T\bm{q}_i - \hat{\mu}_s)^2}$
          \State $\hat{r}_u^s \gets \hat{\mu}_s + \beta_u \hat{\sigma}_s$

        \State $e_{u}^s \gets (\hat{r}_{u}^s - r_{u}^s)$

        \State $\bm{q} \in \mathbb{R}^f$  $\gets 0, \;\bm{v} \in \mathbb{R}^f$  $\gets 0$
        \For {each item $i \in \mathcal{S}$ }
          \State $\bm{q} \gets \bm{q} + \bm{q}_i$
          \State $\bm{v} \gets \bm{v} + (\bm{p}_u^T\bm{q}_i)\bm{q}_i$
        \EndFor
        \State $\nabla p_u \gets \frac{\bm{q}}{|\mathcal{S}|} 
          + \frac{\beta_u \bm{v}}{\hat{\sigma_s}|\mathcal{S}|}
          - \frac{\beta_u \mu_s \bm{q}}{\hat{\sigma_s} |\mathcal{S}|}$

        \State $\nabla q \gets \frac{2e_{u}^s p_u}{\mathcal{S}}$
        \For {each item $i \in \mathcal{S}$ }
          \State $t \gets 1 + \frac{\beta_u p_u^Tq_i}{\hat{\sigma}_s} 
            - \frac{\beta_u \mu_s}{\hat{\sigma}_s}  $
            \State $q_i \gets q_i - \eta(t \nabla q + 2\lambda q_i)$
            \Comment{Update item's latent representation}
        \EndFor

        \State $p_u \gets p_u - \eta(2 e_{u}^s \nabla p_u + 2 \lambda p_u )$
        \Comment{Update user's latent representation}
        
        \State $\beta_u \gets \beta_u - \eta(2 e_{u}^s \hat{\sigma_s} +
        2\lambda\beta_u)$ \Comment{Update $\beta_u$}

      \EndFor
      \State $iter \gets iter + 1$
    \EndWhile
    \EndProcedure
  \end{algorithmic}
\end{algorithm}

\subsection{Model learning}\label{model_learn}
The parameters of the models that estimate item- and set-level ratings are the
user and item latent vectors ($p_u$ and $q_i$), in the case of \ES method
the users' weights on extremal subsets ($W$) and in the case of the \VO method
the user's pickiness level ($\beta_u$). These parameters are estimated using the
user-supplied set-level ratings by minimizing a square error loss function given
by

\begin{equation} \label{eq_rmse}
  \mathcal{L}_{rmse}(\Theta) \equiv \sum_{u \in U} \sum_{\substack{s \in
  \mathcal{R}_{u}^s}} (\hat{r}_{u}^s(\Theta) - r_{u}^s)^2,
\end{equation}

\noindent where $\Theta$ represents model parameters, $U$ represents all the users, $\mathcal{R}_{u}^s$ contains all the sets rated 
by user $u$, $r_{u}^s$ is the original rating of user $u$ on set $\mathcal{S}$ 
and $\hat{r}_{u}^s$ is the estimated rating of user $u$ on set $\mathcal{S}$.

To control model complexity, we add regularization of the model parameters
thereby leading to an optimization process of the following form

\begin{equation} \label{eq_obj}
  \minimize_{\Theta} \mathcal{L}_{rmse}(\Theta)  + \lambda (||\Theta||^2),
\end{equation}

\noindent where $\lambda$ is the regularization parameter. The L2-regularization is added to 
reduce the model complexity thereby improving its generalizability. 
This optimization problem can be solved by Stochastic Gradient Descent
(SGD)~\cite{r22} algorithm.

Note that for the \ES model, we
need to solve this optimization problem with linear and non-negative constraints
on user weights $\bm{w_u}^T$. If we know the users' and the items' latent factors
then the user weights can be determined by solving the Equation~\ref{eq_obj} as
a constraint quadratic programming~\cite{boyd2004convex} for each user. 
We can determine a user's weights by solving multiple quadratic programs, each
corresponding to a different extremal subset having the highest weight, and
selecting the solution that has lowest RMSE over the user's sets.
Hence, for \ES we solve for $W$ and $\{\bm{p_u}, \bm{q_i}\}$ alternately at each SGD iteration. 
In \ES, the minimum weight of the extremal subset having highest contribution towards
ratings on sets, i.e., $c$, can be specified in the range [0,1].
Also, in the \VO method 
we add a fixed constant, i.e., $\epsilon$ in [0, 1], to computed $\sigma$ for robustness.
Algorithms~\ref{alg:alg-lfs-arm},~\ref{alg:alg-lfs-esarm}~and~\ref{alg:alg-lfs-voarm}
show the steps used to learn the \ARM,
\ES and \VO models, respectively.

If we also have ratings for the individual items, then we can incorporate these ratings into the model estimation process by treating each item as a set of size one. 
Note that, when we do not have set-level ratings but only have item-level ratings, then the proposed methods reduce to MF as we need to estimate a single item-level rating to estimate the set-level rating.

%% file: experiments.tex

\subsection{Dataset}\label{EVALDATA}
We evaluated the proposed methods on two datasets: (i) the dataset analyzed in
Section~\ref{dataset}, which will be referred to as \MLREALSETS, and (ii) a set
of synthetically generated datasets that
allow us to assess how well
the optimization algorithms can estimate accurate models and how their
accuracy depends on various data characteristics.

The synthetic datasets were derived from the \ML
20M
dataset\footnote{https://grouplens.org/datasets/movielens/20m/}~\cite{harper2016movielens}  which
contains 20 million ratings from approximately 229,060 users on 26,779 movies.
For experiment purposes, we
created a synthetic low-rank matrix of rank 5 as follows. 
We started by generating two matrices
$A\in\mathbb{R}^{n\times k}$ and $B\in\mathbb{R}^{m\times k}$, where $n$ is
number of users, $m$ is number of items and $k = 5$, whose values are
uniformly distributed at random in $[0, 1]$. We then computed the singular value
decomposition of these matrices to obtain $A=U_A\Sigma_A V_A^T$ and $B=U_B\Sigma_B
V_B^T$. We then let $P=\alpha U_A$, $Q=\alpha U_B$ and $R = PQ^T$. Thus, the final
rank $k$ matrix $R$ is obtained as the product of two randomly generated rank $k$
matrices whose columns are orthogonal. Note that the parameter $\alpha$ was
determined empirically in order to produce ratings in the range of $[-10, 10]$.

Since we know the complete synthetic low-rank matrix we can generate the rating
corresponding to an observed user-item pair in the real dataset from the complete
rating matrix.
We randomly selected 1000 users without replacement from the dataset and for
each user we created sets containing five movies. 
The movies in a user's set
were selected at random without replacement from the movies rated by that user.
For each user, we created at least $k$ such sets of movies, where $k \in [40,
60, 80, 100, 140]$.
We generated rating for a user on a set by 
following two approaches: 
\begin{enumerate}[(i)]  
  \item \ES-based rating: For each user, we chose one of the
extremal subsets at random and used that to generate ratings for all the sets.
The set is assigned an average of the user's ratings on the items in the
chosen extremal subset of the items in the set. 

  \item  \VO-based rating: For each user, 
we chose the user's level of pickiness (the $\beta_u$ parameter) at random from the
range [-2.0, 2.0]. The set is assigned an average of the user's ratings on the items in the set,
and also we offset this rating by adding a quantity computed by scaling the
standard deviation of ratings in the set by the randomly chosen user's level of
pickiness.
\end{enumerate}

For all these datasets, we added random $\mathcal{N}(0, 0.1)$ Gaussian noise while computing
ratings at both the item and set-level for the users. 
For each approach, we generated 15 different synthetic datasets, each
by varying the user-item latent factors and the users' pickiness.

\subsection{Evaluation methodology}
To evaluate the performance of the proposed methods we divided the available
set-level ratings for each user into training, validation and test splits by randomly
selecting five set-level ratings for each of the validation and test splits. 
The validation split was used for model selection.
In order to assess the performance of the methods for item
recommendations, we used a test set that contained for each user the items that
were not present in the user's sets (i.e., these were absent from the training,
test, and validation splits) but were present in the original user-item rating
matrix used to generate the sets.
We used Root Mean Square Error (RMSE) to measure the accuracy of the rating prediction
over items and sets. 

\subsubsection{Comparison methods}\label{comp_methods}
In addition to the evaluation of the proposed methods, i.e., ARM, \ES 
and \VO, we also present the results for the following methods:
\begin{enumerate}[(i)]
  \item SetAvg: This personalized method predicts a user's ratings on items and sets as the average of the
    user's ratings on sets. The rating of user $u$ on set $\mathcal{S}$ is given
    by
    \begin{equation}
      \begin{split}
        \hat{r}_{u}^s &= \frac{1}{|\mathcal{Q}_u|} \sum_{k \in \mathcal{Q}_u} r_{uk}, 
      \end{split}
    \end{equation}
  \noindent where $\mathcal{Q}_u$ represents all the sets rated by user $u$. The
  rating of user $u$ on item $i$ is given by
  \begin{equation}
    \begin{split}
      \hat{r}_{ui} &= \frac{1}{|\mathcal{Q}_u|} \sum_{k \in \mathcal{Q}_u} r_{uk},
    \end{split}
  \end{equation}
\noindent where $\mathcal{Q}_u$ represents all the sets rated by user $u$.

  \item Item average: This non-personalized method estimates the rating for an item as the average of the ratings
provided by the users on the item. The rating
$\hat{r_i}$ for an item $i$ is given by
\begin{equation}
  \begin{split}
    \hat{r}_{i} &= \frac{1}{|\mathcal{U}_i|}\sum_{u \in \mathcal{U}_i} r_{ui}, 
  \end{split}
\end{equation}
\noindent where $\mathcal{U}_i$ denotes the set of users who have rated item
$i$.

  \item UserMeanSub: This non-personalized method estimates the rating for an item as the sum of average rating on
sets and average of user mean subtracted item ratings.
The rating
$\hat{r_i}$ for an item $i$ is given by
\begin{equation}
  \begin{split}
    \hat{r}_{i} &= \mu_s + \frac{1}{|\mathcal{U}_i|}\sum_{u \in \mathcal{U}_i}
    \big(r_{ui} - \frac{1}{|\mathcal{I}_u|}\sum_{k \in \mathcal{I}_u}r_{uk}\big),
  \end{split}
\end{equation} 
\noindent where $\mu_s$ is the average of the ratings on all the sets, $\mathcal{I}_u$
represents the set of items rated by user $u$.

\item \MFSET: This personalized method assumes that a user's ratings on the items in a set are equal to the rating provided by the user on the set. It uses these item-level ratings to estimate the user and the item latent factors by using the MF method. The rating of user $u$ on item $i$ is given by 
\begin{equation} \label{mfset_eq_dot_prod}
  \begin{split}
    \hat{r}_{ui} = \bm{p_u}^T\bm{q_i},
  \end{split}
\end{equation}
\noindent where the vector $\bm{p_u} \in \mathbb{R}^f$ denotes the $f$ dimensional user's latent representation and similarly for item $i$, the vector 
$\bm{q_i} \in \mathbb{R}^f$ represents the $f$ dimensional item's latent
representation.

\item \MFOPT: This method uses the actual users' ratings on the items in the set to estimate the user and the item latent factors by using the MF method. Note that given that \MFOPT uses the actual item-level ratings, its performance will be better than the other methods that rely only
    on set-level ratings. As such, \MFOPT's performance can be used to assess the \emph{opportunity cost} associated with using set-level ratings over using the corresponding item-level ratings. 

\end{enumerate}

In practice, 
a significant proportion of the ratings provided by users on items
depends on factors that are associated with either users or items, and do not
depend on interactions between the users and the items. For example, some users
have a tendency to rate higher than others, and some items tend to receive higher
ratings than others. For the real set-level rating dataset, that we obtained from
\ML users, we model these tendencies by estimating user- and
item-biases~\cite{koren2009matrix} as part of the model learning.

\subsection{Model selection}
We performed grid search to tune the dimensions of the latent factors and
regularization hyper-parameters for the latent factors. 
We searched
for regularization weights ($\lambda$) in the range [0.001, 0.01, 0.1, 1, 10],
$\epsilon$ in the range [0.1, 0.25, 0.5, 1] and $c$ in the range [0, 0.25, 0.50,
0.75, 0.90] for both the synthetic and the real
datasets.
We searched for the dimension of latent factors ($f$) in the range [1, 5, 10,
15, 25, 50, 75, 100] for real datasets, and used 5 as the dimension
of latent factors for synthetic datasets.
The final
parameters were selected based on the performance on the validation split.

%% file: results.tex

The experimental evaluation of the various methods that we developed is done in three
phases. First, we investigated how well the proposed models can explain the users' ratings over sets in the dataset we obtained from a subset of Movielens users
(described in Section~\ref{dataset}). Second, we evaluated the performance of the methods using the synthetically
generated datasets in order to assess how well the underlying optimization algorithms
can recover the underlying data generation models and achieve good prediction
performance at either the set- or item-level. Note that unless otherwise specified,  we report the average of RMSEs of all the
synthetic datasets as the final RMSE values for each rank and proposed approach. Finally, we evaluated the prediction performance achieved by the proposed methods at both the set- or item-level in the real dataset.

\subsection{Agreement of set-rating models with the observed data} \label{fit_rating_model}
In order to determine how well the proposed models can explain the ratings that the
users in our dataset provided, we performed the following analysis. We selected
sets with standard deviation of at least 0.5, and included only those users who
have rated at least 20 such sets. This left us with 17,552 sets rated by 493
users.

\begin{figure}[tb]
  \centerline{\includegraphics[scale=0.8]{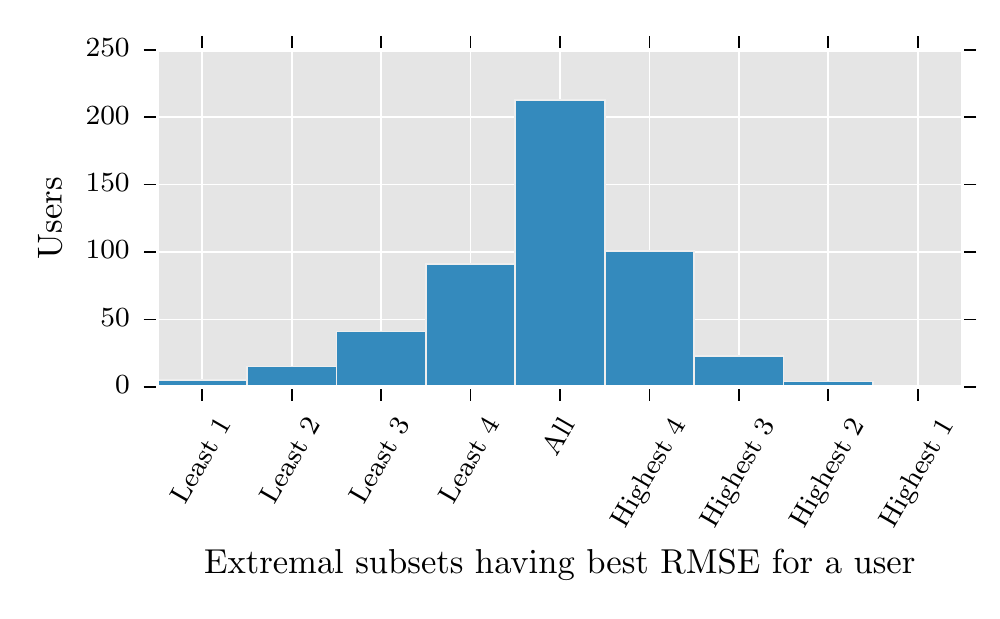}}
  \caption{The number of users for which their pickiness behavior is explained
  by the corresponding least- and highest-rated subsets of
items.}
  \label{fig:extremal}
\end{figure}

To study the \ES model, for each set rated
by a user we created all the possible subsets having either $k$ lowest or $k$
highest rated items for all the possible values of $k$ $\in$ [1, 5], i.e., nine
extremal subsets.
We computed the error between the average rating of items in the extremal
subsets and the rating provided by a user on a set. Similarly, we computed the
error over the remaining sets for a user and selected that subset among the nine
extremal subsets corresponding to which the user has lowest Root Mean Square
Error (RMSE) for all the sets.
Figure~\ref{fig:extremal} shows the number of users and their corresponding
extremal subset that obtained lowest RMSE for their sets. As can be seen in the
figure, there are certain users for whom the
lowest RMSE on sets corresponds to either $k$ lowest or $k$ highest rated items in a set,
where $k < 5$. This indicates that while providing a rating to a set of items, the user may get
influenced more by a subset of the items in a set rather than  all the items in
the set.

\begin{figure}[tb]
  \centerline{\includegraphics[scale=0.9]{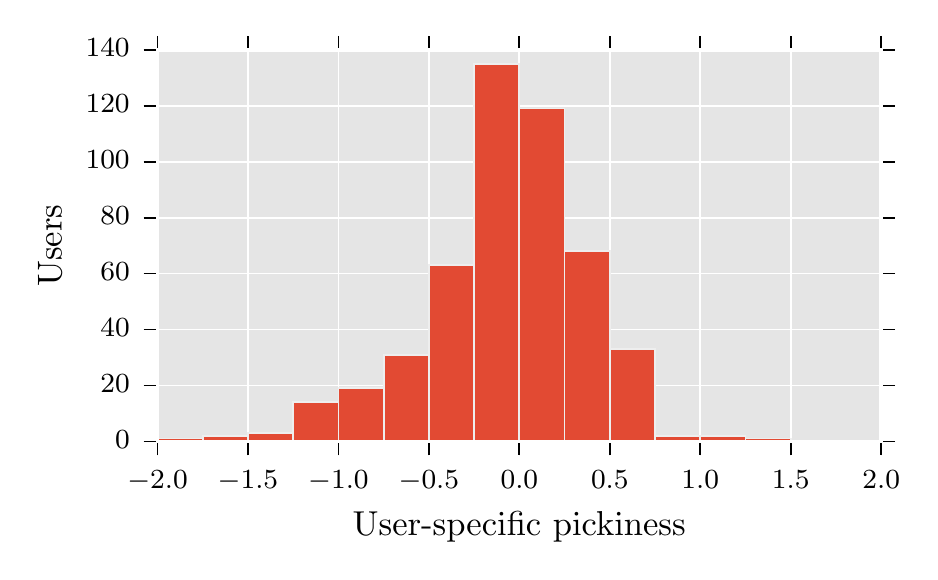}}
  \caption{The number of users and their computed level of pickiness.}
  \label{fig:pickiHist}
\end{figure}

Further, to investigate \VO model, we computed the user's level of pickiness ($\beta_u$) as 
\begin{equation} \label{pickinessEq}
  \begin{split}
    \beta_u &= \frac{1}{n_s}\sum_{s = 1}^{n_s} \frac{r_{u}^s - \mu_s} {\sigma_s},
  \end{split}
\end{equation}
where $n_s$ is the number of sets rated by user $u$, $r_{u}^s$ denotes the rating
provided by user $u$ on set $s$, $\mu_s$ is the mean rating of the items in set $s$
and $\sigma_s$ is the standard deviation of the ratings of the items in set $s$.
Figure~\ref{fig:pickiHist} shows the histogram of the  users' level of pickiness. As
can be seen from the figure,  certain users tend to under- or over-rate sets with
high standard deviation. We conducted a $t$-test on the magnitude of values of pickiness between the set of users that under-rate sets and the set of users that over-rate sets. We found the behavior of users under- and over-rating sets to be statistically significant ($p$-value < 0.001 using $t$-test).
Furthermore, we observe that more users (268) tend to under-rate sets than over-rate them (224).

Additionally, we computed how well the above rating models, i.e., \ES and \VO,
compare against the \ARM model where a user rates
a set as the average of the ratings that he/she gives to the set's items.
We used the user-specific pickiness determined in above analysis for the \ES and
the \VO models to estimate a user's rating on a set.
Table~\ref{table:fit_results} shows the RMSE of the estimated ratings according to different models
and as can be seen in the table both the \ES and the \VO give a better fit to the
real data than \ARM, thereby suggesting that modeling users' level of  pickiness could lead
to better estimates.

\begin{table}[t]
  \centering
  \caption{Fit of different rating models on the data}
  \label{table:fit_results}
  \begin{threeparttable}
  \def\arraystretch{1.5}
  \begin{tabular}{lccc}
  \hline
      &\multicolumn{1}{c}{\centering \ARM} 
      &\multicolumn{1}{c}{\centering \ES} 
      &\multicolumn{1}{c}{\centering \VO} \\ 
  \hline
  RMSE    & 0.597  & 0.509  & 0.521 \\
  \hline
  \end{tabular}
  \end{threeparttable}
\end{table}


\subsection{Performance on the synthetic datasets}

\subsubsection{Accuracy of set- and item-level predictions}
\begin{figure}[t]
  \centerline{\includegraphics[scale=0.57]{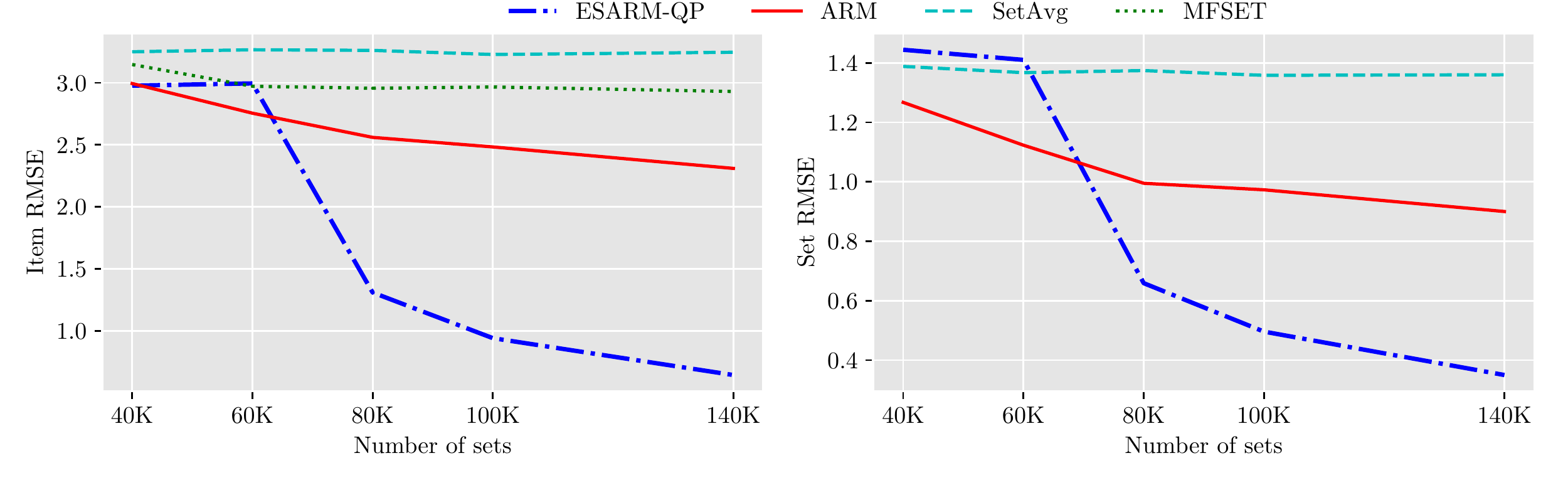}}
  \caption{The average RMSE obtained by the proposed methods on \ES-based datasets with different number of sets.}
  \label{fig:es_sets_sz}
\end{figure}

\begin{figure}[t]
  \centerline{\includegraphics[scale=0.57]{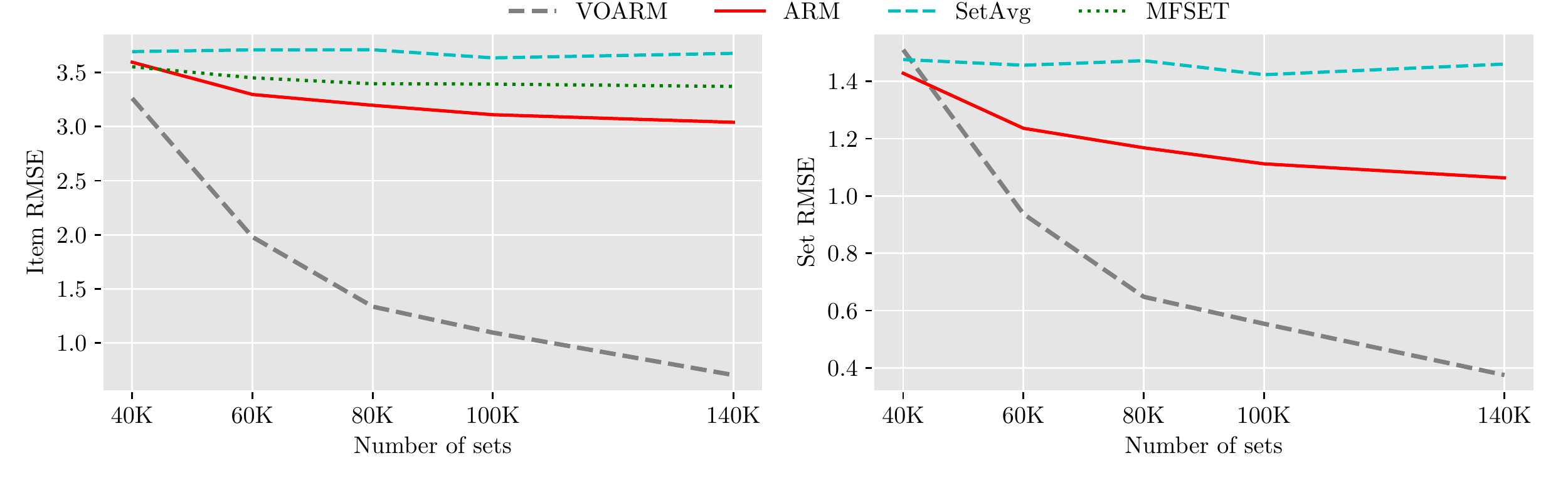}}
  \caption{The average RMSE obtained by the proposed methods on \VO-based datasets with different number of sets.}
  \label{fig:vo_sets_sz}
\end{figure}

We investigated the performance of the proposed methods for both item- and
set-level predictions on the synthetic datasets.
In addition to the performance of each method on its corresponding dataset,
we also show the performance of the \ARM and \SETAVG methods in Figures~\ref{fig:es_sets_sz}~and~\ref{fig:vo_sets_sz}. 

Figure~\ref{fig:es_sets_sz} shows that \ES outperforms all other methods for 
both set- and item-level predictions for datasets with a large number of sets.
However, for datasets with fewer sets, \ARM  outperforms \ES and \SETAVG for the set- and item-level predictions. Additionally, \ARM outperforms \MFSET for item-level predictions as well.
Figure~\ref{fig:vo_sets_sz} shows that \VO outperforms all other methods for both
set- and item-level predictions. 
Unlike \ES, \VO performs better than other methods even for the
case when we have fewer sets, and this suggests that \ES needs a
larger number of sets than \VO to recover the underlying characteristics of the
data. 
Note that even though both \ARM and \MFSET cannot model the underlying pickiness characteristics of the datasets, the former does considerably better than the latter. We believe that this is due to the fact that \ARM's model, which assumes that the average of the set's item-level ratings is equal to that of the set's rating is significantly more flexible than \MFSET's model, which assumes that both the set and all of its items have exactly the same rating. This flexibility allows ARM to better model sets in which there is a high variance among the ratings of the set's items. To test this hypothesis, we performed a series of experiments in which we generated sets with progressively more diversely rated items, which showed that the gap between \ARM and \MFSET increased with the diversity of ratings in sets (results not shown). Since \ARM and \MFSET have the same motivation and \ARM outperforms \MFSET method, we will present results for the \ARM method in the remaining section.

\subsubsection{Recovery of underlying characteristics}\label{syn_recov}
\begin{figure}[t]
 \centerline{\includegraphics[scale=0.7]{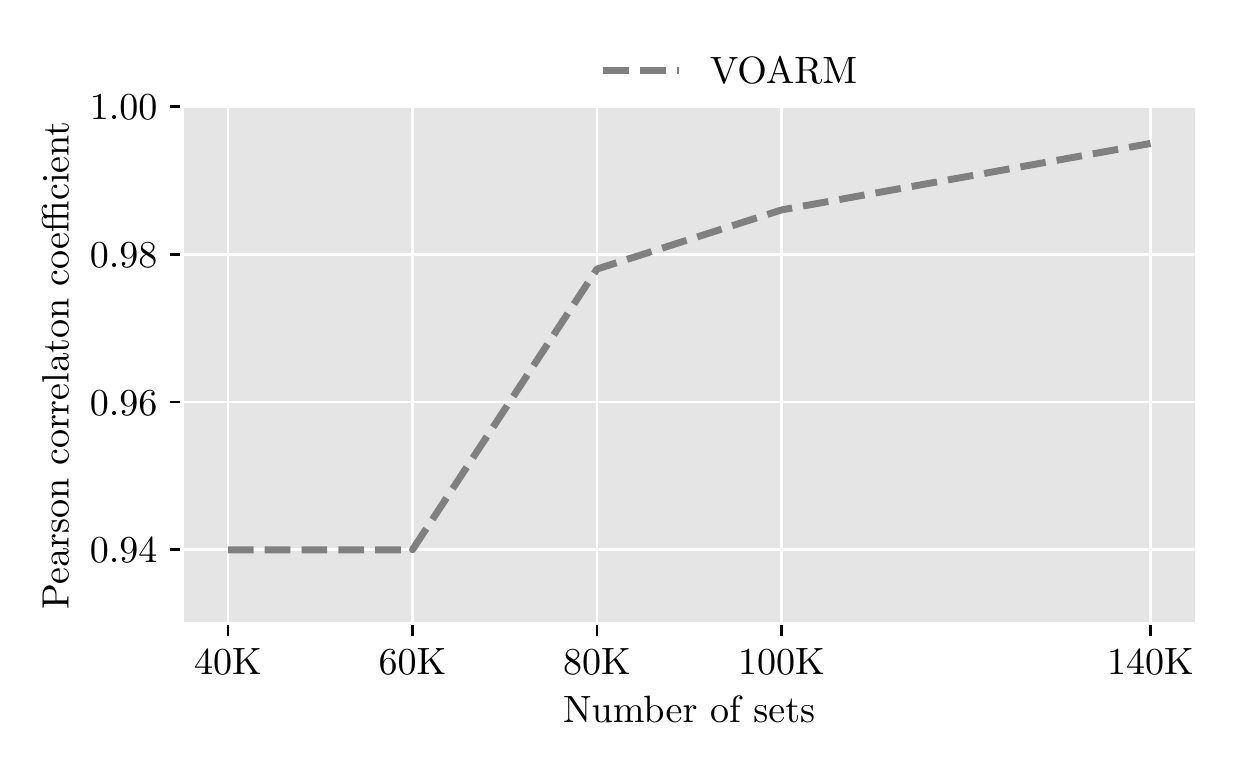}}
  \caption{Pearson correlation coefficients of the actual and the estimated
    parameters that model a user's level of pickiness in the \VO model.}
  \label{fig:vo_pearson}
\end{figure}

\begin{figure}[t]
  \centerline{\includegraphics[scale=0.7]{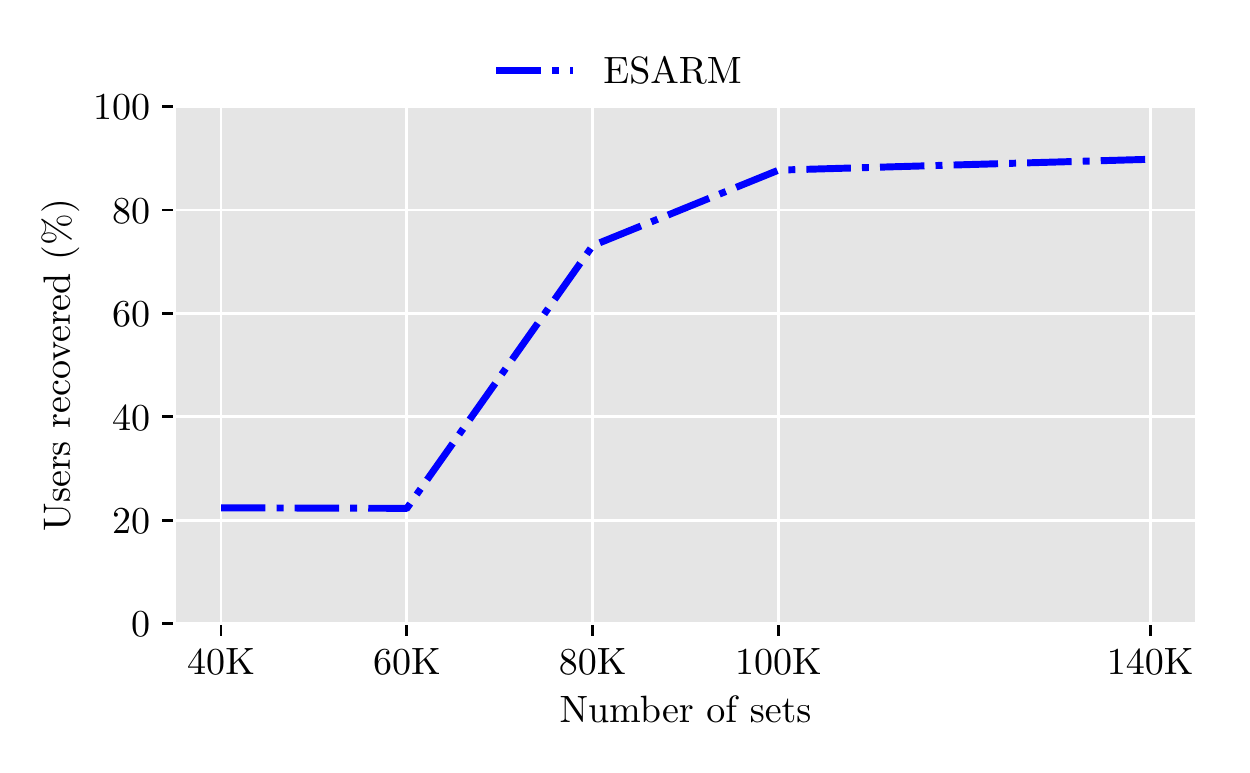}}
  \caption{The percentage of users recovered by \ES, i.e., the users for whom the
    original extremal subset had the highest estimated weight under these models. 
}
  \label{fig:es_sets_recov}
\end{figure}

We examined how well \ES and \VO recover the known underlying characteristics
of the users in the datasets. Figure~\ref{fig:vo_pearson} plots the Pearson correlation coefficient of
the actual and the estimated weights that model the users' level of pickiness in
\VO (i.e., $\beta_u$ parameters). The high values of Pearson correlation
coefficients in the figure
suggests that \VO is able to recover the overall characteristics of the
underlying data. Additionally, this recovery of underlying characteristics
increases with the increase in the number of sets. In order to investigate how well \ES can recover the underlying characteristics,
we computed the fraction of users for whom the extremal subset having the
highest weight ($w_{ui}$) is same as that of the extremal subset used to generate
the rating on sets.   
Figure~\ref{fig:es_sets_recov} shows the percentage of users for whom the extremal subsets are
recovered by \ES. 
As can be seen in the figure, the fraction of users recovered
by \ES increases significantly with the increase in the number of
sets.
The better performance of \ES on the larger
datasets suggests that in order to recover the underlying characteristics of the
data accurately, \ES needs significantly more data than required by
\VO method. Furthermore, for both \ES and \VO methods, we have a low recovery when we have 40K to 60K sets in the dataset and we believe that this low recovery is because the proposed methods, i.e., \ES and \VO, do not have sufficient data to recover the underlying characteristics and once we increase the number of sets, i.e, $\ge$ 80K, we have sufficient data to recover the underlying user-behavior that generated the set-level ratings.

\begin{figure}[t]
  \centerline{\includegraphics[scale=0.57]{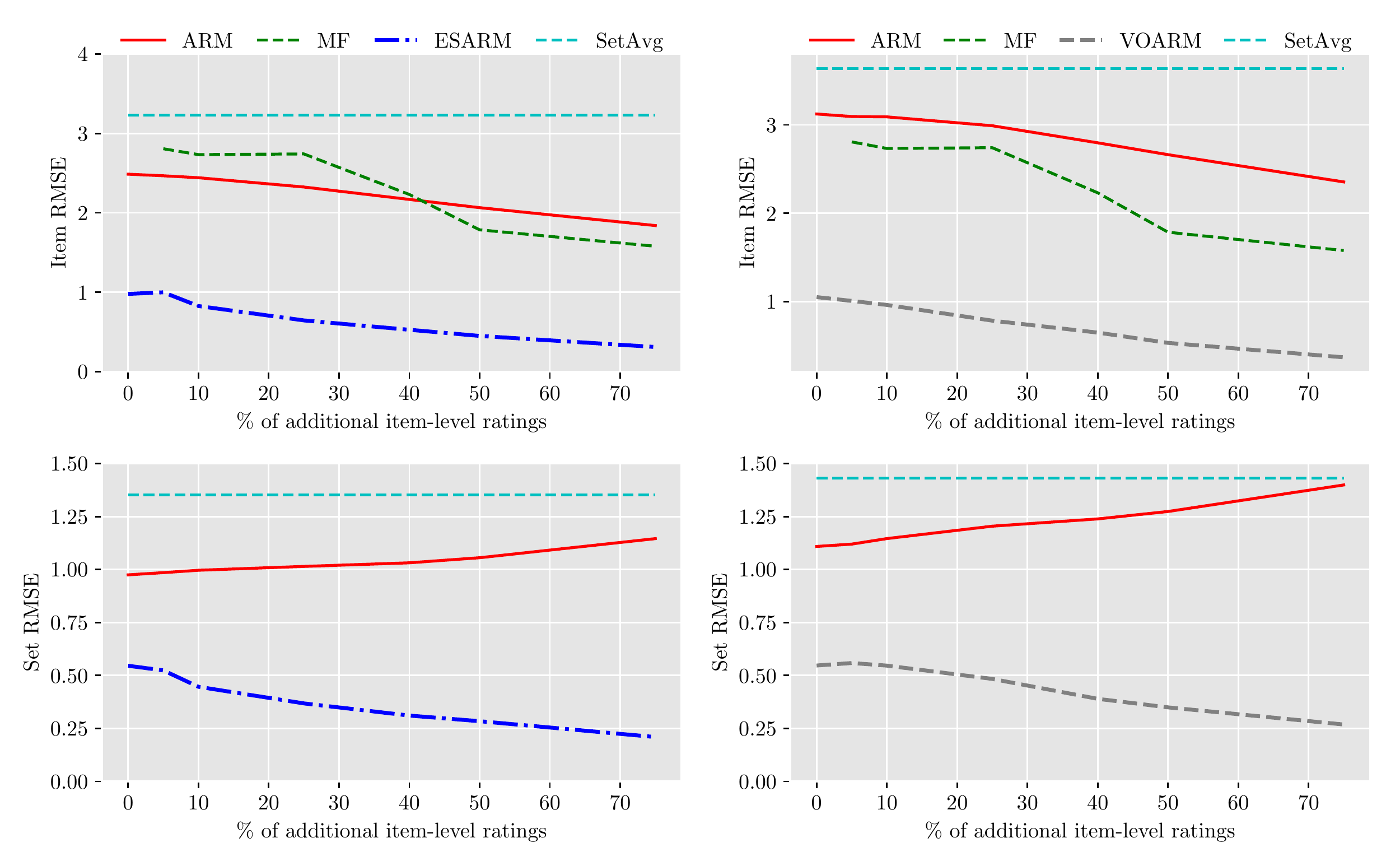}}
  \caption{Effect of adding disjoint item-level ratings for the users in \ES-based (left) and \VO-based
  (right) datasets.}
  \label{fig:syn_add_items}
\end{figure}

\begin{figure}[t]
  \centerline{\includegraphics[scale=0.57]{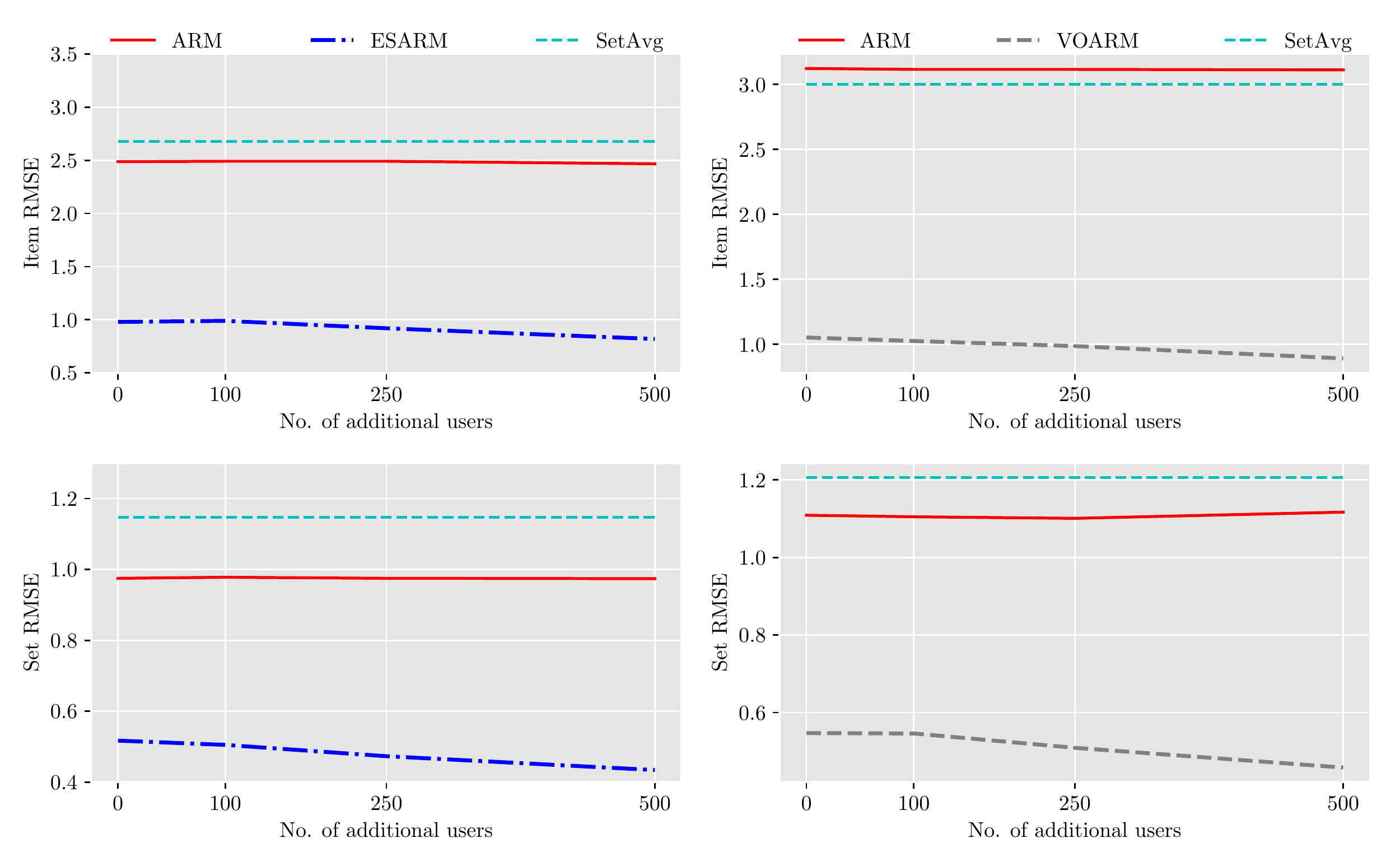}}
  \caption{Effect of adding item-level ratings from additional users in \ES-based (left) and \VO-based
  (right) datasets.}
  \label{fig:syn_add_users}
\end{figure}

\subsubsection{Effect of adding item-level ratings}

In most real-world scenarios, in addition to set-level ratings, we will also have
available ratings on individual items, e.g., users may provide ratings
on music albums and on tracks in the albums. Also, there may exist
some users that are not concerned about keeping their item-level ratings
private.
To assess how well \ES and \VO  can take advantage of such item-level ratings
(when available) we performed three sets of experiments.

In the first experiment, we studied how the availability of additional 
item-level ratings from the users ($U_s$)  that provided set-level ratings affects the performance of the proposed methods. 
To this end, we added in the synthetic datasets a set of item-level ratings for the same set of
users ($U_s$) for which we have approximately $100K$ set-level ratings. The number of item-level ratings was
kept to $k$\% of their set-level ratings, where $k \in [5, 75]$, and the items that were added were disjoint
from those that were part of the sets that they rated. 
In the second experiment, we investigated how the availability of 
item-level ratings from additional users (beyond those that exist in the 
synthetically generated datasets) affect the performance of the proposed 
approaches. We randomly selected 100, 250 and 500 additional users ($U_i$) and added 
a random subset of 50 ratings per user from the items that belong to the sets of users in $U_s$. 
Figures~\ref{fig:syn_add_items}~and~\ref{fig:syn_add_users} shows the
performance achieved by \ES and \VO in these experiments.
Additionally, we used the
matrix factorization (MF) method to estimate the user and item latent
factors by using only the added item-level ratings from the users in  $U_s$.
As can be seen from Figure~\ref{fig:syn_add_items}, as we continue to add item-level ratings for the users in $U_s$, there is an increase
in accuracy of both the set- and item-level predictions for  \ES and
\VO. Both \ES and \VO outperform \ARM  with the
availability of more item-level ratings. For the task of item-level rating
prediction, \ES and \VO even outperform MF,
which is estimated only based on the additional item-level ratings.
Figure~\ref{fig:syn_add_users} shows how the performance of the proposed methods changes when
item-level ratings are available from users in  $U_i$. Similar to the
addition of item-level ratings from users in $U_s$, \ES and \VO
outperform \ARM with the availability of item-level ratings
from users in $U_i$.
The performance of \ARM and SetAvg are significantly lower as both of these methods fail to recover the underlying pickiness characteristics of the dataset and tend to mis-predict many of the item-level ratings.
These results imply that using item-level ratings from the users that provided set-level ratings or from another set of users improves the performance of the proposed methods. 

\begin{table}[bt]
  \centering
  \caption{Average item-level RMSE performance of \ES and \VO for a set of additional users ($U_i$), that have provided only item-level ratings.}
  \label{table:perf_addu}
  \begin{threeparttable}
  \def\arraystretch{1.5}
    \begin{tabular}{@{\hspace{20pt}}l@{\hspace{40pt}}c@{\hspace{20pt}}c@{\hspace{20pt}}}
      \hline
      Type of ratings & \ES &\VO \\
      \hline
      Item-level ($U_i$) &  2.860 & 2.860\\
      Set-level ($U_s$) + item-level ($U_i$) &  \underline{1.811} & \underline{1.866} \\
      \hline
    \end{tabular}
    \begin{tablenotes}[para,flushleft]
    \footnotesize
        $U_i$ is the set of additional 500 users that have provided only item-level ratings.
      $U_s$ is the set of users that have provided set-level ratings.
      Item-level ($U_i$) denotes the item-level ratings of users in $U_i$. 
      Set-level ($U_s$) denotes the set-level ratings of users in $U_s$.
    \end{tablenotes}
  \end{threeparttable}
\end{table}

In the third experiment, we investigate if using set-level ratings
from one set of users ($U_s$) can improve the item-level predictions for another set of
users ($U_i$) for whom we have only item-level ratings. We randomly selected 500 additional users
($U_i$) and added a random subset of 50  ratings per user from the items that belong to
the sets rated by existing users ($U_s$). 
Table~\ref{table:perf_addu} shows the performance of item-level predictions for 
users in $U_i$  and performance on these users  after using set-level ratings from users in $U_s$.
As can be seen in the table, the performance of item-level predictions for users
in $U_i$ improves significantly after using set-level ratings from existing users in $U_s$.
These results suggest that using both item- and set-level ratings
not only lead to better item recommendations for users ($U_s$) with set-level ratings but also for those additional users ($U_i$) who have provided item-level ratings.

\subsection{Performance on the Movielens-based real dataset}

\begin{table}[t]
  \centering
  \caption{The RMSE performance of the proposed methods with user- and
  item-biases on \MLREALSETS dataset.}
  \label{table:real_wbias_results}
  \begin{threeparttable}
  \def\arraystretch{1.5}
  \begin{tabular}{@{\hspace{10pt}}l@{\hspace{40pt}}c@{\hspace{40pt}}c@{\hspace{10pt}}}
  \hline
    Method & Item & Set\\
  \hline
  SetAvg  & 0.976  & 0.630\\
  \ARM    & \underline{0.971} & 0.624\\
  \ES  & 0.979 & 0.631\\
  \VO     & 0.973 & \underline{0.623}\\ 
  \hline
  \end{tabular}
  \end{threeparttable}
\end{table}

Our final experiment used the proposed approaches (\ARM, \ES, and \VO) to
estimate both set- and item-level rating prediction models using the real
set-level rating dataset that we obtained from \ML users. 

\subsubsection{Accuracy of set- and item- level predictions}
Table~\ref{table:real_wbias_results} shows results for the case when we have
only set-level ratings. 
As can be seen in the table, \ARM outperforms the remaining methods for item-level 
predictions. However, \VO performs somewhat better than \ARM for set-level predictions. The better performance
of \ARM for item-level predictions is not surprising as most of the sets in the dataset
are rated close to the average of the ratings on items in sets. Also, as seen in
our analysis in Section~\ref{syn_recov}, \ES needs a large number of sets
in order to accurately recover the users' extremal subsets. 
The difference between the predictions of different models was found to be statistically significant  ($p$-value $\le$ 0.02 using $t$-test). 
Table~\ref{table:pairwise_method_comp} shows the percentage of the item-level predictions for whom a proposed approach performs better than the other approaches. As can be seen in the table, \ARM and \VO performs better than other methods for the majority of the item-level predictions. 
In addition, \VO performs better than \ARM for the majority of the item-level predictions. The lower RMSE of \ARM for item-level predictions and better performance of \VO for the majority of the item-level predictions suggest that there are few item-level predictions where \VO's error is significantly higher than that of ARM.
In
Section~\ref{picky_users_analysis}, we will investigate the performance of the
proposed methods independently for picky and non-picky users.  

\begin{table}[bt]
    \caption{Percentage of item-level predictions where method X performs better
    than method Y.}
    \label{table:pairwise_method_comp}
    \centering
    \begin{tabular}{c|cccc}
         \hline
         \backslashbox{Method X}{Method Y} & SetAvg & \ARM & \ES & \VO \\
         \hline
         SetAvg & - & 49.56 & 53.74 & 46.41  \\
         \ARM & 50.44 & - & 51.01 & 49.85 \\
         \ES & 46.26 & 48.99 & - & 45.54 \\
         \VO & 53.59 & 50.15 & 54.46 & - \\
         \hline
    \end{tabular}
\end{table}

\subsubsection{Effect of adding item-level ratings}
In addition, we assessed how
well the proposed methods can take advantage of additional item-level
ratings. 
In the first experiment, we added $k$\% of the users' set-level ratings, where $k
\in [10, 75]$, as additional item-level ratings and the items that were added 
were disjoint from those that were part of the sets that they rated. 
In the second experiment, we added ratings from 100, 250 and 500 additional users (beyond
those that have participated in the survey), and these users have provided on an
average 20,000 ratings for the items that belong to the existing users' sets.
In the third experiment, we studied if using set-level ratings from existing users can improve recommendations for additional users that provided only item-level ratings. To this end, we randomly selected 500 additional users and added a random subset of 10 ratings per user from the items that belong to the sets rated by existing users.

\begin{figure}[bt]
  \centerline{\includegraphics[scale=0.8]{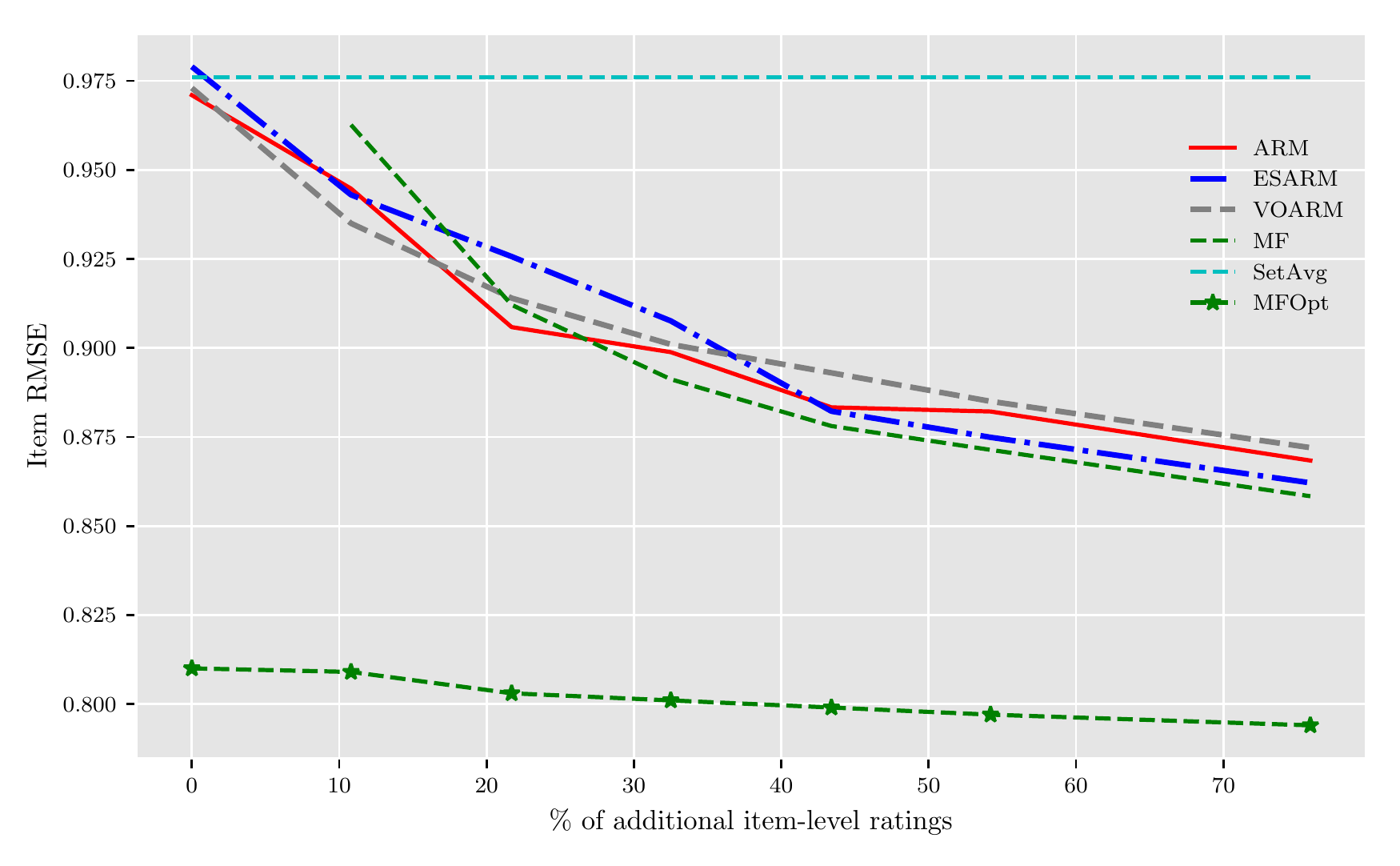}}
  \caption{Effect of  adding item-level ratings from the same set of users in the real dataset.}
  \label{fig:real_add_items_wbias}
\end{figure}

Figure~\ref{fig:real_add_items_wbias} shows the results obtained for the first experiment (i.e., adding item-level ratings for the same set of users for which we have set-level ratings). These results show that, with the exception of \SETAVG, the performance of all set-based methods improves as item-level ratings are used and these improvements increase with the percentage of item-level ratings that are used.

\begin{figure}[bt]
  \centerline{\includegraphics[scale=0.57]{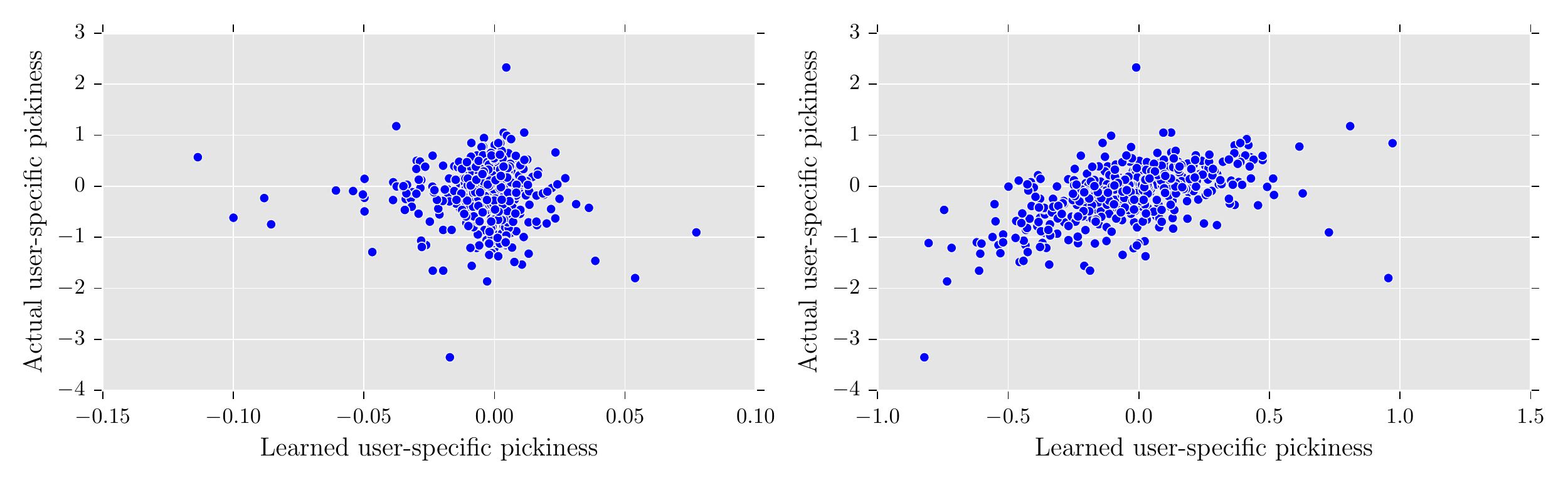}}
  \caption{
    Scatter plots of the user's original level of pickiness computed from real
    data and the pickiness estimated by \VO from set-level ratings (left), and
    after including 30\% of item-level ratings (right).  
  }
  \label{fig:real_add_items_recov}
\end{figure}

Besides the set-based methods, Figure~\ref{fig:real_add_items_wbias} also reports the performance of MF, which uses only the added item-level ratings and the performance of \MFOPT, which in addition to the added item-level ratings it uses the actual item-level ratings of the items in the sets (see discussion in Section~\ref{comp_methods}). 
These results show that when the number of item-level ratings is small ($<30$\%), MF does not do as well as the set-based methods; however, when there is a sufficiently large number of item-level ratings, MF outperforms the set-based methods, indicating that the set-based methods can improve the recommendations when we have set-level ratings and do not have sufficient item-level ratings to use MF with high accuracy in the recommender system.
Additionally, when MF outperforms set-based methods and when set-based methods outperform MF, we found the differences between the predictions from MF and the set-based methods to be statistically significant ($p$-value < 0.01 using $t$-test).
Figure~\ref{fig:real_add_items_recov} plots the estimated weights that model a
user's level of pickiness in \VO against the user's level of pickiness, i.e.,
$\beta_u$, computed
from the data in Section~\ref{fit_rating_model}. As can be seen in the figure, to some
extend \VO is
able to recover the user' level of pickiness after addition of few item-level
ratings.
Also, the difference between the performance of the proposed methods and \MFOPT is reduced as we continue to add more item-level ratings.

\begin{figure}[bt]
  \centerline{\includegraphics[scale=0.8]{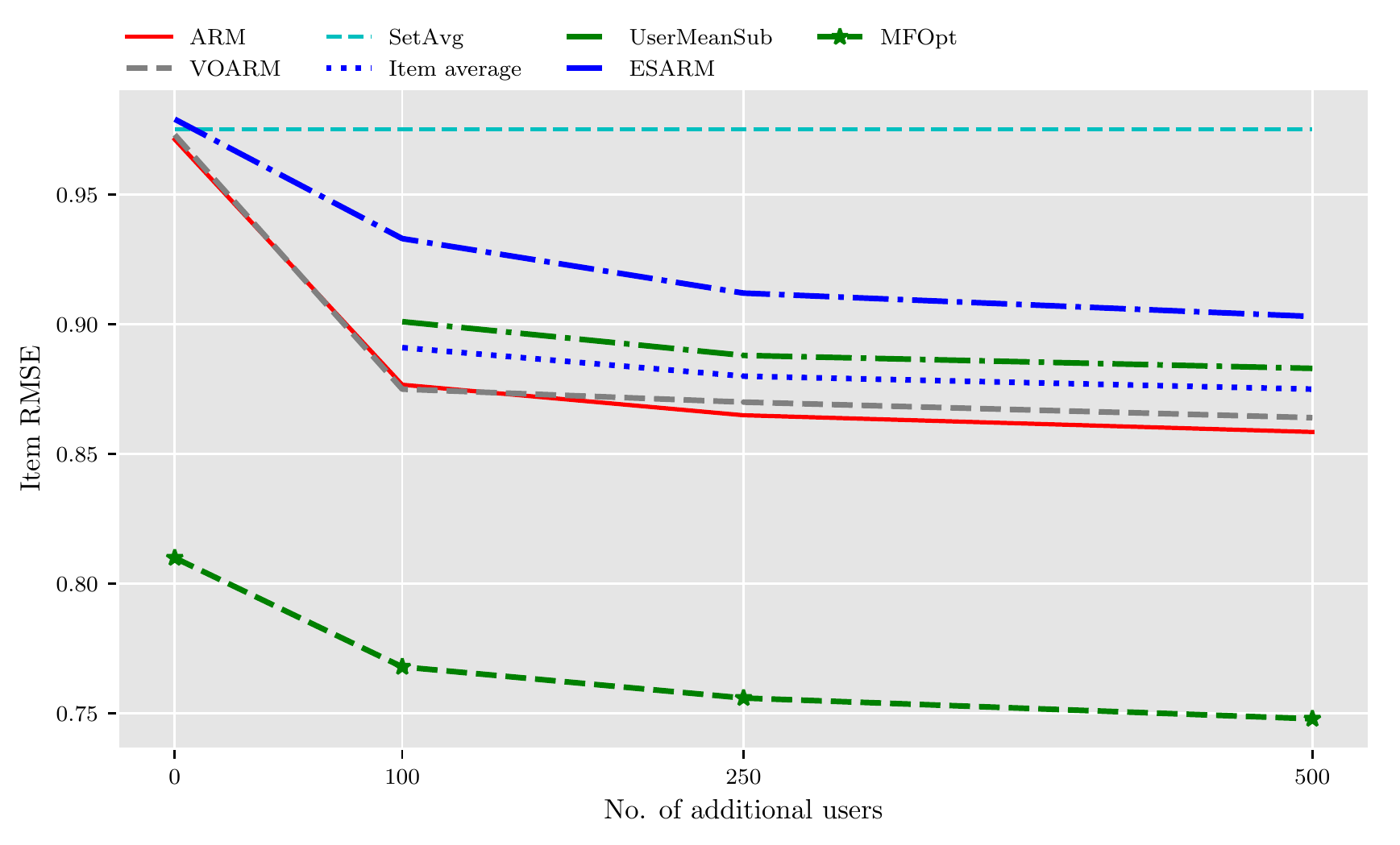}}
  \caption{Effect of modeling biases and adding item-level ratings from disjoint set of users in the real dataset.}
  \label{fig:real_add_users_wbias}
\end{figure}

In the second experiment, we examined the case when we have item-level ratings from the
additional users. 
In addition to estimating ratings from the proposed methods, we estimated the ratings at item-level from the two non-personalized methods, i.e., Item average and UserMeanSub, as described in Section~\ref{comp_methods}. 
Figure~\ref{fig:real_add_users_wbias}  shows the results for these non-personalized methods along with that of the proposed methods. As can be seen in the figure, \VO and \ARM outperform the non-personalized methods and since \ES needs a large number of sets to recover the users' extremal subsets, it does not outperform the non-personalized methods. Furthermore, the performance of the proposed methods continue to improve with the availability of more item-level ratings from additional users. Additionally, the difference between the performance of the proposed methods and MFOpt is reduced as we add more item-level ratings from additional users.

\begin{table}[bt]
  \centering
  \caption{RMSE  for item-level predictions for
  additional users, that have provided only the item-level ratings.}
  \label{table:perf_addu_real}
  \begin{threeparttable}
  \def\arraystretch{1.5}

  \begin{tabular}{@{\hspace{10pt}}l@{\hspace{40pt}}c}
    \hline
    Method & Item-level RMSE \\
    \hline
    MF  & 1.003 \\
    \ARM & \underline{0.978} \\
    \ES & 1.043 \\
    \VO & 1.033 \\
    \hline
  \end{tabular}
  \end{threeparttable}
\end{table}

In the third experiment, we investigated if using set-level ratings from existing users can
improve the item-level predictions for additional users who have provided
ratings only at item-level. 
Table~\ref{table:perf_addu_real} shows the performance of item-level predictions for additional users
after using set-level ratings from the existing users and also shows the
performance of MF method after using only the additional item-level ratings.
As can be seen in the table, \ARM outperforms MF for item-level predictions after
using set-level ratings from existing users. However, \ES and \VO do not
perform better than MF for the additional users. 
Similar to our results on synthetic datasets, it is promising that using
item-level ratings from the additional users and set-level ratings from the
existing users improves the performance not only for latter but also for those
additional users who have provided only item-level ratings.

\subsubsection{Accuracy of item-level predictions for picky
users}\label{picky_users_analysis}
\begin{table}[bt]
  \centering
  \caption{The item-level RMSE of the proposed methods on different subset of users using
  only set-level ratings and after including additional item-level ratings.}
  \label{table:perf_picky_subsets}
  \begin{threeparttable}
  \def\arraystretch{1.5}
    \begin{tabular}{@{\hspace{20pt}}l@{\hspace{20pt}}c@{\hspace{8pt}}c@{\hspace{20pt}}c@{\hspace{8pt}}c@{\hspace{20pt}}}
      \hline
      & \multicolumn{2}{c@{\hspace{20pt}}}{Set only} & \multicolumn{2}{c@{\hspace{20pt}}}{+Items}\\ 
      
      Method & \unonpicky & \upicky & \unonpicky & \upicky \\
      \hline
      \ARM & \underline{0.915} & 1.089 & \underline{0.879} & 0.975 \\
      \ES & 0.922 & 1.103 & 0.898 & \underline{0.923} \\
      \VO & 0.921 & \underline{1.085} & 0.892 & 0.932 \\
      \hline
    \end{tabular}
  \begin{tablenotes}[para,flushleft]
    \footnotesize
    The ``Set only'' column denotes the results of the models that were
    estimated using only set-level ratings. The ``+Items'' column show the
    results of the models that were estimated using the sets of ``Set only'' and
    also some additional ratings on a different set of items from the same users
    that provided the set-level ratings.
    \upicky refers to the users who have rated at least 20 sets and have a high
    level of pickiness, i.e., $|\beta_u| > 0.5$, in real dataset, and \unonpicky represents
    the remaining users.  
  \end{tablenotes}
  \end{threeparttable}
\end{table}

Even though
\ARM performs better than remaining methods for item-level predictions, we investigated how
well do \ARM, \ES and \VO perform for item-level predictions for the users who have rated at least 20 sets and
have a high level of pickiness, i.e., $|\beta_u| > 0.5$. We found 374 users in the
dataset that were non-picky (\unonpicky) and 135 users that were having a higher level of
pickiness (\upicky). 
Table~\ref{table:perf_picky_subsets} shows the performance of the proposed
methods for item-level predictions using set-level ratings and after
including 30\% of additional item-level ratings on both \upicky and
\unonpicky. As can be seen in the table, for set-level ratings
\VO performs somewhat better than \ARM on \upicky and
after including additional item-level ratings  both \ES and \VO
outperform \ARM on \upicky.

The overall consistency of the results between the synthetically generated and
the real dataset suggests that \VO and to some extend \ES 
are able to capture the tendency that some users have to consistently under- or
over-rate  diverse sets of items.

\subsection{Summary}
In this work, we investigated two questions related to using set-level ratings in recommender systems. First, how users' item-level ratings relate to the ratings that they provide on a set of items. Second, how collaborative-filtering-based methods can take advantage of such set-level ratings towards making item-level rating predictions. Based on the set of experiments that were presented, we can make the following overall observations:
\begin{enumerate}
    \item The \ES and \VO set-rating models can explain the ratings provided by users on sets of items better than the \ARM model, with \ES doing slightly better than \VO (Section~\ref{fit_rating_model}).
    
    \item The proposed models can use both item- and set-level ratings to improve recommendations not only for users who provided ratings on sets but also for users with only item-level ratings.
    
    \item When the ratings follow \ES and \VO rating models, the proposed models can recover the underlying characteristics in the data and are resilient to noise in these ratings (Section~\ref{syn_recov}).
    
    \item Users with high level of pickiness, i.e., $|\beta_u| > 0.5$, \VO recovers the underlying characteristics in the data better than \ARM and after including additional item-level ratings \ES outperforms both \ARM and \VO in terms of recovery of underlying characteristics in the data (Section~\ref{picky_users_analysis}).
\end{enumerate}

%% file: conclusion.tex

In this work, we studied how users' ratings on sets of items relate to their
ratings on the sets' individual items. 
We collected ratings from active users of \ML on sets of movies and based
on our analysis we developed collaborative filtering-based models that try to explicitly model the
users' behavior in providing the ratings on sets of items.
Through extensive experiments on synthetic and real data, we showed that the
proposed methods can model the users' behavior as seen in the real data and
predict the users' ratings on individual items.

For future work, it will be interesting to study how do the performance of the proposed
approaches vary with the different number of items in sets and how do they
perform when instead of having a fixed number of items in sets, the sets contain
a varied number of items in sets. Furthermore, one can use ratings on sets of items to 
generate a ranked list of items by optimizing a ranking loss~\cite{rendlebpr09} over the ratings on 
sets of items.
Additionally, in our work, we have used the matrix factorization approach to estimate item-level ratings and alternatively, we can also use other recently proposed approaches (e.g., deep learning-based methods~\cite{sedhain2015autorec,he2017neural}) to estimate item-level ratings.
Also, the performance of the model could be improved by modeling
temporal effects on the ratings and by using side-information like genres or
other movie metadata. Finally, it will be interesting to investigate if similar to the diversity of
ratings in the set there exists other properties at item- or set-level
that can affect a user's ratings on sets of items.